\documentclass[11pt]{article}
\usepackage{fullpage}
\usepackage{enumerate} 

\usepackage[ascii]{inputenc}
\usepackage[T1]{fontenc}
\usepackage[english]{babel}
\usepackage{times}
\usepackage{epsfig}
\usepackage{amsmath}
\usepackage{amssymb}
\usepackage{amsfonts}
\usepackage{amstext}
\usepackage{latexsym}
\usepackage{graphicx, subfigure}
\usepackage{longtable}

 \newcommand{\qed}{\hfill $\square$  \smallbreak}

 \newtheorem{theorem}{Theorem}[section]
  \newtheorem{property}{Property}[section]
 \newtheorem{claim}[theorem]{Claim}
  \newenvironment{proof}
 {\noindent{\bf Proof:}}
 {\qed}

\newcommand{\al}{
\textbf{a}
}
\newcommand{\alp}{
\textbf{a.}
}

\newcounter{smallitemizec}

\begin{document}

\title{Exploration of Periodically Varying Graphs\thanks{This work was partially supported by ARC, ANR Project SHAMAN,  by COST Action 295
 DYNAMO, and by NSERC.} }

\date{ }
 \author{
 Paola Flocchini\thanks{SITE, University of Ottawa, Ottawa, Canada.
 {\tt flocchin@site.uottawa.ca} }
 \and
 Bernard Mans \thanks{Macquarie University, Sydney, Australia.
{\tt bmans@science.mq.edu.au}
}
 \and
 Nicola Santoro \thanks{
School of Computer Science, Carleton
 University,
 Ottawa,  Canada.
 {\tt santoro@scs.carleton.ca}
}
}
\maketitle

\begin{abstract}
We study the computability and complexity of the
{\em exploration problem} in a   class of highly dynamic graphs: 
  {\em periodically varying} (PV)  graphs, where  the edges exist 
only at some (unknown) times  defined by the periodic movements
 of  carriers.
These graphs naturally model  highly dynamic
   infrastructure-less networks  such as public transports with fixed timetables, low earth orbiting (LEO) satellite systems, security guards' tours, etc.

We establish necessary conditions
for the problem to be solved. 
We also derive lower bounds on the amount of time required 
in general, as well as for the PV graphs defined by
 restricted classes  of  carriers movements: 
 simple routes, and circular routes.

We then prove  that the limitations on computability and
complexity we have established  are indeed tight.
In fact we  prove that all necessary conditions are also sufficient
and all lower bounds on costs are tight.
We do so constructively presenting two
worst case optimal solution 
algorithms, one for anonymous systems, and
one for those with distinct nodes ids. 
An added benefit is that the algorithms
are rather simple.

\end{abstract}


 \pagenumbering{arabic}

 %%%%%%%%%%%%%%%%%%%%%%%%%%%%%%%%%%%%%%
 \section{Introduction}
  %%%%%%%%%%%%%%%%%%%%%%%%%%%%%%%%%%%%%%

{\em Graph  exploration}  
is a classical fundamental problem
extensively studied since its initial formulation  in $1951$ by Shannon
\cite{Sh51}. It has   various applications in
different areas, e.g,
finding a path through a maze, or
searching a computer network using a mobile software agent.  In
these cases, the environment to be explored is usually modelled as a
(di)graph, where a single entity (called agent or 
robot) starting at  a node of the graph, has to visit
all the nodes and terminate within finite time.
Different instances of the problem exist  depending on a variety of factors, including 
whether the nodes of the graph are
labelled with unique identifiers or are anonymous,
 the amount of memory with which the exploring  agent is endowed, 
 the amount of a priori knowledge available about the structure of the graph
 (e.g., it is acyclic) etc.  (e.g., see 
\cite{AlbHe00,AwBS99,BenFRSV98,CFIKP08,DeP99,DP04}). 
In spite of their differences, all these investigations have something in common:
they all assume that the graph to be explored is  {\em connected}. 

 The connectivity  assumption unfortunately does not hold for
the new generation of networked
environments 
that are highly dynamic and evolving in time. 
In these infrastructure-less networks,
end-to-end multi-hop paths may not exist, and
 it is actually possible that,
 at every instant of time, the network is disconnected.
 However,  communication routes may be available 
 through time and mobility, and 
  not only basic tasks like routing, 
but complex communication and computation services could still be performed.
 See in this regard the ample literature (mostly from the  application/engineering community)
 on  these highly dynamic systems, 
  variously
 called
 {\em delay tolerant}, {\em disruption tolerant}, 
  {\em challenged}, and {\em opportunistic} networks (e.g., \cite{BuiFJ03,BurGJN03,Guk07,JacMR09,JaiFP04,LiW09b,OdW05,SpPR05,ZhangxKLTZ07,Zhangz056}).
Almost all  the existing work in this area  focuses on the {\em routing} problem. 
In spite of the large amount of literature,
no work exists on {\em exploration} of such networks,
with the noticeable exception  of  the study  of exploration  by  random walks   \cite{AvKZ08}.

The highly dynamic features of  these networks can  be described by means  of {\em time-varying}  graphs, that is graphs where links between 
nodes exist only at some times (a priori unknown to the algorithm designer);
thus, for example, the static graph defined by the set of edges existing at a given time might not be connected.
Our research  interest is on  the deterministic  {\em exploration} of time-varying graphs,
on the computability and complexity aspects of this problem.

%
%  %------------------------------------------------------------
%  \subsection{Periodically Varying Graphs}
% %------------------------------------------------------------
 
 In this paper, we start the investigation  focusing  on a particular class  of time-varying graphs:
 the {\em periodically varying graphs} (PV graphs), where
  the edges of the graphs are defined by the periodic movements
 of some mobile entities, called carriers.
 This class models naturally   infrastructure-less networks where mobile entities have fixed routes 
that they traverse regularly. Examples of such common settings are public transports with fixed timetables, low earth orbiting (LEO) satellite systems, security guards' tours, etc.;
these networks have been investigated 
in the application/engineering community, with
 respect to routing and to the design of carriers' routes
 (e.g., see   \cite{Guk07,LiW09b,ZhangxKLTZ07}).

We view the system as composed of $n$ sites and $k$ carriers, each periodically moving among
a subset of the sites. The routes of the  carriers define the edges of the time-varying graph:
a directed edge exists from node $u$ to node $v$ at time $t$ only if there is a carrier that
in its route moves from $u$ to $v$ at time $t$. 
If all routes have the same period the system is called {\em homogeneous}, 
otherwise it is called {\em heterogeneous}.
In the system enters an explorer agent \al that can ride with any carrier 
along its route,  and it can switch to any carrier it meets while riding.
Exploring  a PV-graph is the process
of   \al visiting  all the nodes  and exiting the
system within finite time.
 We study the computability and the complexity 
of   the exploration problem of PV-graphs, {\em PVG-Exploration}.

We first  investigate the computability of {\em PVG-Exploration}
and establish necessary conditions for the problem to
be solvable.
%relationship between 
% availability of a priori knowledge by \al about the system    and
% (deterministic) solvability of
%       the problem.
We prove that
 in {\em anonymous} systems (i.e., the nodes have no identifiers) 
 exploration is unsolvable if  the agent has 
no knowledge of (an upper bound on) the size of the largest route;
if the nodes have {\em distinct ids}, we show that
 either $n$ or an upper-bound on the
       system period  must be known for the problem to be solvable. 
       
These necessary conditions  
for anonymous systems, summarized in the table below,  
hold even if the routes are {\em homogeneous} (i.e., have the same length),
the agent has unlimited memory and 
knows $k$\footnote{if anonymous, even if they know $n$}.

  \begin{center}
\begin{tabular}{|c|c|c|}
\hline
                   ANONYMOUS   & &\\
                   \hline
                   \hline
                      {\tt  Knowledge}  &{\tt Solution}       & ({\tt Even if})\\
 \hline
   \hline
   \hline
 (bound on)    $p$  unknown          &   {\em impossible}  &   $n,k$ known; homogeneous   \\
%         &    &   unbounded memory   \\
    \hline
    \hline
(bound on)    $p$ known           &   {\em possible}
% ( {\sc Post-Order})
 &   $n,k$ unknown; heterogeneous \\
%         &    &   $\Theta(\log p + k \log k)$ bits\\
%(bound on)    $p$ known          &   {\em possible} &   ($n,k$ unknown; hetero; \\
%         &   ( {\sc Post-Order})  &  $\Theta(\log p + k \log k)$ bits)\\
       \hline
       \end{tabular}
       \end{center}

\begin{center}
\begin {tabular}{|c|c|c| }
\hline
                DISTINCT IDS & &    \\
   \hline
   \hline

           {\tt  Knowledge}  &{\tt Solution}       & ({\tt Even if})\\
 \hline
   \hline
 $n$ and (bound on)  $p$  unknown  &   {\em impossible} &   $k$ known; homogeneous \\
%        &    &  unbounded memory  \\
\hline
\hline
$n$  known            &    {\em possible}
% ({\sc Guessing})
 & $p,k$ unknown; heterogeneous \\
        &    &    $O(n \log n)$ bits \\
    \hline
(bound on)    $p$ known           &   {\em possible}
%  ( {\sc Post-Order})
   &   $n,k$ unknown; heterogeneous \\
          &    &       $O(\log p + k \log k)$ bits\\
       \hline
       \end{tabular}

\end{center}

We then consider the complexity  
 of {\em PVG-Exploration} and establish  lower bounds on the number of moves.
 We prove that  in general
$ \Omega(k p) $ moves are necessary for homogeneous systems
and $ \Omega(k p^2)$ for heterogeneous ones,
where $p$  is the length of   the longest route. 
This lower bound holds 
 even if
    \al knows $n, k, p$, and has unlimited
    memory.
 Notice that the parameter $p$ in the above lower bounds can
be  arbitrarily large since  the same node can appear in a route 
arbitrarily many times.
A natural question is whether the lower bounds   do change 
imposing restrictions on the  ``redundancy"  of the routes.
To investigate the impact of the routes' structure on the complexity of the problem,
 we consider  PV-graphs where all the routes are  {\em simple}, that is,   do not contain
self-loops nor multiple edges.  We show that the same type of 
lower bound holds also for this class;
 in fact,
 we establish  $\Omega(k n^2)$ lower bound  for    homogeneous and
   $\Omega(k n^4)$ lower bound for   heterogeneous systems with simple routes,
We then further restrict  each route to be
{\em circular}, that is   an edge
appears in a route at most once. Even in this case,
the general lower bound holds; in fact  we prove lower bounds of
 $\Omega(k n)$ moves for homogeneous  and   $\Omega(k n^2)$ 
 for heterogeneous systems with circular routes.
 Interestingly these lower bounds hold even if \al has full knowledge of
 the entire PV graph, and has  unlimited
    memory.
  We then prove  that the limitations on computability and
complexity established so far,  are indeed tight.
In fact we  prove that all necessary conditions are also sufficient
and all lower bounds on costs are tight.
We do so constructively presenting two worst case optimal solution algorithms,
one  for anonymous systems and one for those with ids.
In the case of {\em anonymous} systems, the algorithm 
solves the problem 
without requiring  any knowledge of $n$ or $k$;
in fact it only uses the necessary knowledge of an upper bound 
 $B\geq p$  on
   the size of the longest route. 
     The number of moves  is  $O(k B)$
for homogeneous and    $O(k B^2)$ for heterogeneous
   systems.
      The  cost    depends on the accuracy of the 
upperbound $B$ on   $p$.
It is sufficient that the upper bound $B$  is linear in $p$ for the algorithm to
be {\em optimal}.
 In the case of systems {\em with ids}, the algorithm 
solves the problem 
without requiring  any knowledge of $p$ or $k$;
in fact it only uses the necessary knowledge of  
 $n$. 
     The number of moves is  $O(k p)$ and  $O(k p^2)$ matching the lower bound.

  \begin{center}
\begin{tabular}{|c||c|c|}
\hline
                    & {\tt System} & \\
                   \hline
                   \hline
                      {\tt  Routes}  &{\em Homogeneous}       & {\em Heterogeneous}\\

   \hline
   \hline
{\em Arbitrary}         &   $\Theta(k p)$  &   $\Theta(k p^2)$   \\
    \hline
{\em Simple}         &  $\Theta(k n^2)$  &   $\Theta(k n^4)$ \\
       \hline
{\em Circular}         &  $\Theta(k n)$  &   $\Theta(k n^2)$  \\
       \hline
       \end{tabular} \label{bounds}
       \end{center}

An added benefit is that the algorithms
are rather simple and use a limited amount of memory.

Several long proofs are in the Appendix.

 %%%%%%%%%%%%%%%%%%%%%%%%%%%%%%%%%%%%%%
 \section{Model and Terminology}
  %%%%%%%%%%%%%%%%%%%%%%%%%%%%%%%%%%%%%%

 %------------------------------------------------------------
  \subsection{Periodically Varying Graphs}
 %------------------------------------------------------------

 The system is composed of a  set $S$ of  {\em sites};
 depending on whether the sites have unique ids or no identifiers, 
 the system will be said to be {\em with ids} or {\em anonymous},
 respectively.  In the system operates 
 a set   $C$ of  mobile entities called {\em carriers} moving among the sites; $|C|= k\leq n = |S|$. 
 Each carrier
 $c$ has a unique identifier $id(c)$ and  an ordered sequence of sites
  $\pi(c)=<x_0,x_1,\ldots,x_{p(c)-1}>$, $x_i \in S$, called {\em route}; for any integer $j$ we will 
 denote by $\pi(c)[j]$  the component $x_i$ of the route where $i= j \mod p(c)$,
 and $p(c)$ will be called the {\em period}
of $\pi(c)$.
% $\pi(c)[j] = x_{ j \mod p(c)}$.
A carrier $c\in C$ moves cyclically along its {\em route} $\pi(c)$:
  at time $t$, $c$ will move from $\pi(c)[t]$ to $\pi(c)[t+1]$ where
the indices are taken modulo $p(c)$. In the following, $x_0$ will be called
the {\em starting} site of $c$, and the set $S(c) = \{ x_0,x_1,\ldots,x_{p(c)-1} \}$, 
will be called the {\em domain} of $c$;
clearly $|S(c)|\leq p(c)$.

Each route $\pi(c)=<x_0,x_1,\ldots,x_{p(c)-1}>$ defines a
directed  edge-labelled multigraph $\vec{G}(c) = (S(c),\vec{E}(c))$,
where  $\vec{E}(c) =  \{(x_i,x_{i+1}, i), 0\leq i<p(c)\}$ and the operations on
the indices are modulo $p(c)$.
 If $(x,y,t$ mod $p(c))\in \vec{E}(c)$, we 
shall say that $c$ {\em activates} the edge $(x,y)$ at time
$t$.
A site $z\in S$ is the {\em meeting point}  (or {\em connection}) of carriers $a$ and $b$ at time $t$
if  $\pi(a)[t] =\pi(b)[t]=z$; that is, there exist  sites $x$ and $y$ such that,  
at time $t-1$,
 $a$  activates the edge $(x,z)$ and  $b$  activates the edge $(y,z)$.
%;
%that is there is an edge from $x_i$ to $x_{i+1}$ labelled  $i+1$
%for $0\leq i<p(c)$.
%%Let $\sigma(c)$ be the circular sequence associated to $\pi(c)=<x_0,x_1,\ldots,x_{p(c)-1}>$
%%(i.e., defined by   {\tt next}$(x_i) = x_{i+1}$ where
%%the indices are taken modulo $p(c)$).
%%A {\em repetition} in $\sigma(c)$ is a subsequence $\alpha$ such that
%%$\sigma(c) = \alpha\alpha\beta$, for some (possibly empty) $\beta$.
%%A route $\pi(c)$  is {\em repetition-free}  if there are no repetitions in $\sigma(c)$.
A route  $\pi(c)=<x_0,x_1,\ldots,x_{p(c)-1}>$  is {\em simple} if 
 $\vec{G}(c)$  does not contain
 self loops nor multiple edges; that is $x_i\neq x_{i+1}$, for  $0\leq i<p(c)\}$, and
if $(x,y,i), (x,y,j)\in \vec{E}(c) $ then $i=j$.
 A simple route $\pi(c)$  is {\em irredundant} (or {\em cyclic} if   $\vec{G}(c)$ is
 either a simple cycle  or a virtual cycle (i.e., a simple traversal of a tree).

We shall denote by $R =\{\pi(c) : c\in C\}$ the set of all routes and by
 $p(R)$ = Max$\{p(c) : c\in C\}$ the maximum {\em period} of the routes in $R$.
When no ambiguity arises, we will denote $p(R)$ simply as $p$. 
 The set $R$ defines  a directed edge-labelled multigraph $\vec{G}_R=(S,\vec{E})$,
 where $\vec{E} = \cup_{c\in C} \vec{E}(c)$,
called  {\em periodically varying graph} (or, shortly, {\em PV graph}).
%    \subseteq S \times S \times \N$  is defined as follows:
%$(x,y,t$ mod $p(c))\in \vec{E}$  iff 
% $\exists c\in C$ such that   
% $(x,y,t$ mod $p(c))\in \vec{E}(c)$. 
 
% {\bf ATTENTION TO THIS: } +++++
%   The number   $|\vec{E}|$  of edges in $\vec{G}_{R}$ will also called
%the {\em span} of $R$, and denoted by $||R||$. +++++

% If $(x,y,t$ mod $p(c))\in \vec{E}(c)$, we 
%shall say that $c$ {\em activates} the edge $(x,y)$ at time
%$t$.
%A site $z\in S$ is the {\em meeting point}  of carriers $a$ and $b$ at time $t$
%if  $\pi(a)[t] =\pi(b)[t]=z$; that is, there exist  sites $x$ and $y$ such that,  
%at time $t-1$,
% $a$  activates the edge $(x,z)$ and  $b$  activates the edge $(y,z)$.

A  {\em concrete walk} (or, simply, {\em walk})
 $\sigma$ 
% of size $m$ 
 in $\vec{G}_R$ is a
 (possibly infinite)  ordered sequence  $\sigma=$<$e_0, e_1, e_2\ldots $> 
   of edges in $\vec{E}$ where  $e_i=(a_i,a_{i+1},i)\in \vec{E}(c_i)$ for some $c_i\in C$, $0\leq i$.
   To each route   $\pi(c)$ in $R$ corresponds an  infinite  concrete walk 
    $\sigma(c)$ in $\vec{G}_R$  where $e_i=(\pi(c)[i],\pi(c)[i+1],i)$ for $i\geq 0$.
   A concrete walk $\sigma$ is
   a {\em concrete cover}   of $\vec{G}_R$ if
   it includes every site: $\cup_{ 0 \leq i \leq |  \sigma|+1} \ \{a_i\} = S$.

     A set of routes $R$ is {\em feasible} if there exists at least one concrete cover of $\vec{G}_R$
     starting from any carrier.
$R$  is {\em homogeneous} if all
 routes have the same period: $\forall a,b\in C,\  p(a)=p(b)$; it is {\em heterogeneous}
 otherwise.
 $R$ is {\em simple}  (resp.  {\em irredundant})
 if  every route  $\pi(c)\in R$ is   simple  (resp.,  irredundant). 
 With an abuse of notation, the above properties of $R$
 will be used also for   $\vec{G}_R$; hence we will accordingly say that 
  $\vec{G}_R$ is feasible (or homogeneous, simple, etc.).

%\item For  $R'\subseteq R$,
%let $||R'||$, called the {\em span} of $R'$,  denote the number of edges in $\vec{G}_{R'}$.
%   We say that $R'\subseteq R$ {\em covers} $S$
%if {\bf to be done}.

 In the following, when no ambiguity arises,
  we will denote $p(R)$ simply as $p$,  $\vec{G}_R$ simply as $\vec{G}$, and 
  $(x,y,t$ mod $p(c))$  simply as $(x,y,t)$.

 %------------------------------------------------------------
  \subsection{Exploring Agent and Traversal}
 %------------------------------------------------------------

In the system is injected an external computational entity
   \al called  {\em exploring  agent};
    the agent is injected at the starting site of some carrier at time $t=0$.
    The only two operations it can perform are: move with a carrier, switch carrier.
Agent       \al can switch from carrier $c$ to carrier $c'$ at site $y$ at time $t$
       % $x$ to $y$ at time $t$ 
        iff it is riding with $c$ at time $t$ and  both $c$ and $c'$ arrive at $y$
 %$c$ that activates the edge $(x,y)$ 
 at time $t$, that is:
% , i.e.  $(x,y,t)\in E(c)$.
%In  other words, an agent  is always with a carrier.
 iff it is riding with $c$ at time $t$ and $\exists x,x' \in S$ such that $ (x,y,t)\in E(c) $ and $(x',y,t)\in E(c')$.

%In the system is injected an  {\em exploring  agent}  \al, an external computational entity that needs
%the carriers to move around: \al can move from $x$ to $y$ at time $t$ iff
%there is a carrier $c$ that activates the edge $(x,y)$ at time $t$, i.e.  $(x,y,t,c)\in E$.
%It is assumed that the agent is injected at the starting site of some carrier
Agent \al does not necessarily know $n$, $k$, nor $\vec{G}$; when at a site $x$ at time $t$,
\al can however determine  the identifier $id(c)$ of each carrier $c$
that arrives at  $x\in S$ at time $t$. 
%Levels of
%{\em  information
%acquisition}:
%Basic: [
%When it meets a carrier $c$ at a site,
%it  finds out the identifier $id(c)$.

The goal of \al is to fully explore the system
within finite time, that is to  {\em visit} every site   and terminate,  exiting the system,
within finite time, regardless of the starting position. 
We will call this problem {\em PVG-Exploration}.

An {\em exploration protocol} ${\cal A}$  is an algorithm that specifies the exploring agent's actions enabling it to  traverse  periodically varying graphs.
More precisely, let 
{\em start}$(\vec{G}_R) =\{\pi(c)[0] : c\in C\}$ be the set of starting sites for a periodically varying graph $\vec{G}_R$,
and let $C(t,x) = \{ \pi(c)[t]=x : c\in C\}$, be the set of
carriers that arrive at  $x\in S$ at time $t\geq 0$. 
Initially, at time $t=0$, \al is at a site $x\in start(\vec{G}_R)$.
If  \al is at node $y$ at time $t\geq 0$,  ${\cal A}$  specifies {\em action}$\in C(t,x)\cup \{$halt$\}$:
if {\em action}$=c\in C(t,x)$,
\al will move with $c$ to $\pi(c)[t+1]$,  traversing the
edge $(x, \pi(c)[t+1], t)$ ; if   {\em action}=halt, \al will terminate the execution and exit the system. 
Hence the  {\em execution} of ${\cal A}$ in $\vec{G}_R$ starting from
injection site $x$  uniquely defines  the
(possibly infinite) concrete walk $\xi(x) =< e_0, e_1, e_2, \dots >$ of the edges traversed by \al
starting from $x$; the walk is infinite if \al never executes  {\em action}=halt, finite
otherwise.

Algorithm ${\cal A}$  {\em solves}  the  {\em PVG-Exploration} of $\vec{G}_R$ if
 $\forall x\in${\em start}$(\vec{G}_R)$, $\xi(x)$ is a finite  concrete cover of $\vec{G}_R$;
that is,
executing ${\cal A}$ in  $\vec{G}_R$, \al visits all sites  of $\vec{G}_R$ and performs
{\em action}=halt,   regardless of the injection site $x\in${\em start}$(\vec{G}_R)$.
%Algorithm  {\em solves}  the  {\em PVG-Exploration} of $\vec{G}_R$ if
%executing ${\cal A}$ in  $\vec{G}_R$, \al visits all sites  of $\vec{G}_R$ and performs
%{\em action}=halt,   regardless of the injection site $x\in${\em start}$(\vec{G}_R)$.
%We will denote by $\T(\vec{G}_R)$
%the problem of \al fully exploring  the periodically varying graph $\vec{G}_R$,
%regardless of the starting position.
%An {\em exploration algorithm} ${\cal A}$  is an algorithm that specifies the agent's specific set of instructions for traversing a periodically varying graph with the requirement that the procedure terminates at some point. More specifically,
%an  exploration algorithm ${\cal A}$ solves $\T(\vec{G}_R)$ if, regardless of the starting
%site, the agent \al executing  ${\cal A}$  visit every site in  $\vec{G}_R$ and terminates in finite time.
%Problem $\T(\vec{G}_R)$ is {\em unsolvable} if there is no deterministic exploration algorithm
%that solves  $\T(\vec{G}_R)$.
%,
%under conditions ${\cal C}$. 
Clearly, we have the following property.

 \begin{property}
     {\em PVG-Exploration} of $\vec{G}_R$ is  possible  only if $R$ is feasible.
\end{property}
%We will consider two cases, depending on whether the alien can survive only if attached to
%a carrier (a {\em parasitic} alien),
%or it can wait at a node for a carrier to arrive ({\em independent} alien).

%A  {\em concrete walk} 
% $\sigma$ of size $m$ in $\vec{G}_R$ is an ordered sequence  $\sigma=$<$e_0,e_1,\ldots, e_{m-1}$> 
%   of edges in $\vec{E}$ where  $e_i=(a_i,a_{i+1},i,c_i)$ for some $c_i\in C$, $0\leq i < m$; 
%   a concrete walk $\sigma$ is
%   a {\em concrete cover}   of $\vec{G}_R$ if
%   it includes every node: $\cup_{0\leq i\leq m} \ \{a_i\} = S$.
%     A set of routes $R$ is {\em feasible} if there exists at least one concrete cover of $\vec{G}_R$.
%    Thus, we have the following:

 \noindent   Hence, in the following,  we will assume that  $R$ is {\em feasible} and restrict  {\em PVG-Exploration}  to the class of feasible periodically varying  graphs.
We will say that problem {\em PVG-Exploration} is {\em unsolvable} (in a class of PV graphs) if there is no deterministic exploration algorithm that solves the problem for all feasible PV graphs (in that class).

%  \begin{itemize}

 The cost measure   is  the number of {\em moves}
 that the exploring agent  \al performs.
% of a {\em exploration algorithm} ${\cal A}$
%  in case of a parasitic agent and  the total amount of {\em time} in case of independent
%agent. Futhermore, a
   Let ${\cal M}(\vec{G}_R)$ denote the number of moves that need to be performed
   in the worst case by \al  to solve  {\em PVG-Exploration} in   feasible $\vec{G}_R$. Given a class ${\cal G}$ of feasible graphs,
   let   ${\cal M}(${\cal G}$)$  be the largest $ {\cal M}(\vec{G}_R)$ over all  $\vec{G}_R\in{\cal G}$;
   and let 
   ${\cal M}_{homo}(n,k)$ (resp.   ${\cal M}_{hetero}(n,k)$)
   denote  the largest $ {\cal M}(\vec{G}_R)$ 
   in the class   of  all  feasible homogeneous  (resp. heterogeneous) PV graphs $\vec{G}_R$ with $n$ sites and
   $k$ carriers.

%The other important cost measure is
% the amount of {\em  memory} required by the exploring agent.

%A  exploration algorithm ${\cal A}$ solves $\T(n,k)$ if it solves  $\T(\vec{G}_R)$
%for all feasible periodically varying graph
%$\vec{G}_R$ with $n$ sites and $k$ carriers.
%Problem $\T(n,k)$ is {\em unsolvable} if there is no deterministic exploration algorithm
%that solves  $\T(n,k)$ in finite time.

  %%%%%%%%%%%%%%%%%%%%%%%%%%%%%%%%%%%%%%
  \section{Computability and Lower Bounds}
   %%%%%%%%%%%%%%%%%%%%%%%%%%%%%%%%%%%%%%
       
   %------------------------------------------------------------
   \subsection{Knowledge and Solvability}
   %------------------------------------------------------------

 The availability of a priori knowledge by \al about the system       has  an immediate impact on the solvability of
       the problem  {\em PVG-Exploration}.
       Consider first {\em  anonymous} systems: the sites are indistinguishable to
       the exploring agent  \alp In this case,  the problem   is  unsolvable if \al
      has no knowledge of  (an upper bound on) the system period.

          \begin{theorem}
     \label{teo:anon-imp-p}
      Let the systems be  {\em anonymous}.   {\em PVG-Exploration}  is  unsolvable if \al
      has no information on (an upper bound on) the system period.
     This result holds even
     if  the systems are restricted to be {\em homogeneous},
      \al   has unlimited
     memory and knows both $n$ and $k$.
         \end{theorem}

         \begin{proof}     
          By contradiction, let ${\cal A}$  solve {\em PVG-Exploration} in all anonymous feasible
          PV graphs without  any information on  (an upper bound on) the system period.        
   Given $n$ and $k$, let       $S=\{x_0, \ldots, x_{n-1}\}$  be a set of $n$ anonymous sites, and 
    let $\pi$ be an arbitrary sequence  of elements of $S$  such that  all sites are included.      
        Consider the homogeneous system  where 
          $k$ carriers have exactly the same route $\pi$
          and let $\vec{G}$ be the corresponding graph. 
          Without loss of generality, let $x_0$ be the starting site.
          Consider now the execution of ${\cal A}$ by \al 
          in $\vec{G}$ starting from $x_0$. Since ${\cal A}$ 
          is correct, 
          the walk $\xi(x_0)$ performed by  \al is a finite  concrete cover; let $m$ be
          its length.
          Furthermore, since all carriers have the same route, 
          $\xi(x_0)$ is a prefix of the infinite walk   $\sigma(c)$, performed by each carrier $c$;
          more precisely it consists of the first $m$ edges of    $\sigma(c)$.
           Let $t_i$ denote the first time when $x_i$ is visited in this execution; without   loss of
          generality, let $t_i < t_{i+1}$, $0\leq i< n-2$. 
          
          Let $\pi^*$ denote the sequence of sites in the order they are
          visited by \al in the walk  $\xi(x_0)$.
             Let $\alpha$ be  the first  $t_{n-2}+ 1$  sites of $\pi^*$, and
             $\beta$  be the next  $m+ 1 - (t_{n-2}+ 1) $ sites (recall,  $m$ is the length of 
           $\xi(x_0)$ and thus $m+1$ is that of $\pi^*$).
       Let $\gamma$ be the sequence obtained from $\beta$ by substituting each occurrence
         of   $x_{n-1}$ with $x_{n-2}$. 
                         
%            The problem was that I had "$x_{n-1}$ with $x_{n}$" instead of
%            $x_{n-1}$ with $x_{n-2}$"         and substituting each occurrence
%         of   $x_{n-1}$ with $x_{n-2}$  as it is now. 

           Consider now the homogeneous system  where all the
          $k$ agents have the same route $\pi' =<\alpha, \gamma, \beta>$,
          and let  $\vec{G'}$ be the corresponding graph.

          The 
          execution of ${\cal A}$ in  $\vec{G'}$ by \al with injection site  $x_0$ 
          results in \al performing a concrete walk 
          $\xi'(x_0)$ which, for the  first $m$ edges, is identical to $\xi(x_0)$  
          except that
          each edge of the form $(x,x_{n-1},t)$ and  $(x_{n-1},x,t)$
          has been replaced by $(x,x_{n-2},t)$ and  $(x_{n-2},x,t)$, respectively.
   Because of anonymity of the nodes, \al will be unable to
         distinguish $x_{n-1}$ and  $x_{n-2}$; furthermore,
         it does not know (an upper bound on) the
         system's period). Thus   \al will be unable to
         distinguish the first $m$ steps of the  two executions; it
         will therefore  stop after $m$ moves also in  $\vec{G'}$. This means
         that \al stops before 
         traversing $\beta$; since  $x_{n-1}$ is neither in $\alpha$ nor in $\gamma$,
         $\xi'(x_0)$ is finite but not a concrete cover of   $\vec{G'}$, contradicting the
          correctness of ${\cal A}$.
                 \end{proof}

   \noindent In other words, in anonymous systems, an upper bound on the system period must be
   available to \al for the problem to be solvable.

Consider now  {\em distinct ids} systems, i.e.  where the sites have distinct identities accessible
         to \al when visiting them; in this case,
       the problem  is  unsolvable if \al      has no  knowledge 
 of neither (an upper bound on) the system period nor of the number of sites.

 \begin{theorem}
 \label{teo:IdImpPn}
Let the sites have {\em distinct ids}.
 {\em PVG-Exploration}  is  unsolvable if \al      has no  information on 
 either (an upper bound on) the system period or of the number of sites.
This result holds even if   the systems are  {\em homogeneous},  and  \al
has unlimited    memory and knows $k$.
\end{theorem}

%---------------------------------
 \begin{proof}    
 %{\bf (of Theorem \ref{teo:IdImpPn})} \\
          By contradiction, let ${\cal A}$  solve {\em PVG-Exploration} in all feasible
          PV graphs  with {\em distinct ids} without any
           information on   either (an upper bound on) the system period
        or on the number of sites.
   Let        $S=\{x_0, \ldots, x_{n-1}\}$  be a set of  $n$ sites with distinct ids, and 
    let $\pi$ be an arbitrary sequence  of elements of $S$  such that  all sites are included.      
        Consider now the homogeneous system  where 
          $k$ carriers have exactly the same route $\pi$
          and let $\vec{G}$ be the corresponding graph. 
          Without loss of generality, let $x_0$ be the starting site.
          
            Consider now the execution of ${\cal A}$ by \al 
          in $\vec{G}$ starting from $x_0$. Since ${\cal A}$ 
          is correct, 
          the walk $\xi(x_0)$ performed by  \al is a finite  concrete cover; let $m$ be
          its length and let $\overline{\pi}$ be the corresponding sequence of nodes.
          Furthermore, since all carriers have the same route, 
          $\xi(x_0)$ is a prefix of the infinite walk   $\sigma(c)$, performed by each carrier $c$;
          more precisely it consists of the first $m$ edges of    $\sigma(c)$.
           Consider now the homogeneous system  
           with $n+1$ sites  $S'=\{x_0, \ldots, x_{n-1}, x_n\}$ where all the
          $k$ agents have exactly the same route $\pi' =<\overline{\pi}  x_n>$,
          and let  $\vec{G'}$ be the corresponding graph. The 
          execution of ${\cal A}$ with injection site  $x_0$ will have \al perform the walk 
          $\xi'(x_0)$ which, for the  first $m$ edges, is identical to $\xi(x_0)$.
          Since \al does not know the number of sites, it will be unable to
         distinguish the change, and
         will therefore  stop after $m$ moves also in  $\vec{G'}$. This means
         that \al stops before 
        visiting $x_n$; that is, 
         $\xi'(x_0)$ is finite but not a concrete cover, contradicting the correctness of ${\cal A}$.
                 \end{proof}

\noindent       In other words,  when the sites have unique ids,  either $n$ or an upper-bound on the
       system period  must be known for the problem to be solvable.

   %------------------------------------------------------------
   \subsection{Lower Bounds on Number of Moves}
   %  in  Arbitrary Periodically Varying Graphs}
   %------------------------------------------------------------

 %------------------------------------------------------------
   \subsubsection{Arbitrary Routes}  \ \\
We will first consider the general case,  where no assumptions are
made on the structure of the system routes, and establish lower bounds
on the  number of  moves  both in homogeneous and heterogeneous systems.

%   Unfortunately, even if  both $n$ and $k$  are  known a priori  there are systems with
%   arbitrarily large periods  that require
%the entire span to be traversed, and this even if the system is homogeneous.
%  In fact, we have the following theorem.

\begin{theorem}
     \label{teo:moveLBarb}
         For any $n, k, p$, with $n\geq 9$, $ \frac{n}{3}\geq k \geq 3$, and  $p \geq \max\{k-1,\lceil \frac{n}{k-1} \rceil \} $,
         there exists  a feasible {\em homogeneous} graph $\vec{G}_R $  with $n$ sites, $k$ carriers and period
        $p$ such that   
%        \begin{center}
  ${\cal M}(\vec{G}_R) \geq
   (k-2) (p+1) \ +   \lfloor \frac{n}{k-1}\rfloor. $
%  \end{center}
    This result holds even if
%    the  system is {\em homogeneous},
    \al knows $ \vec{G}_R, k$ and $ p$, and has unlimited
    memory.
  \end{theorem}

\begin{figure}[tbh]	
      \label{fig:moveLBarb}
 	\centering
     \includegraphics[width=0.8\textwidth]{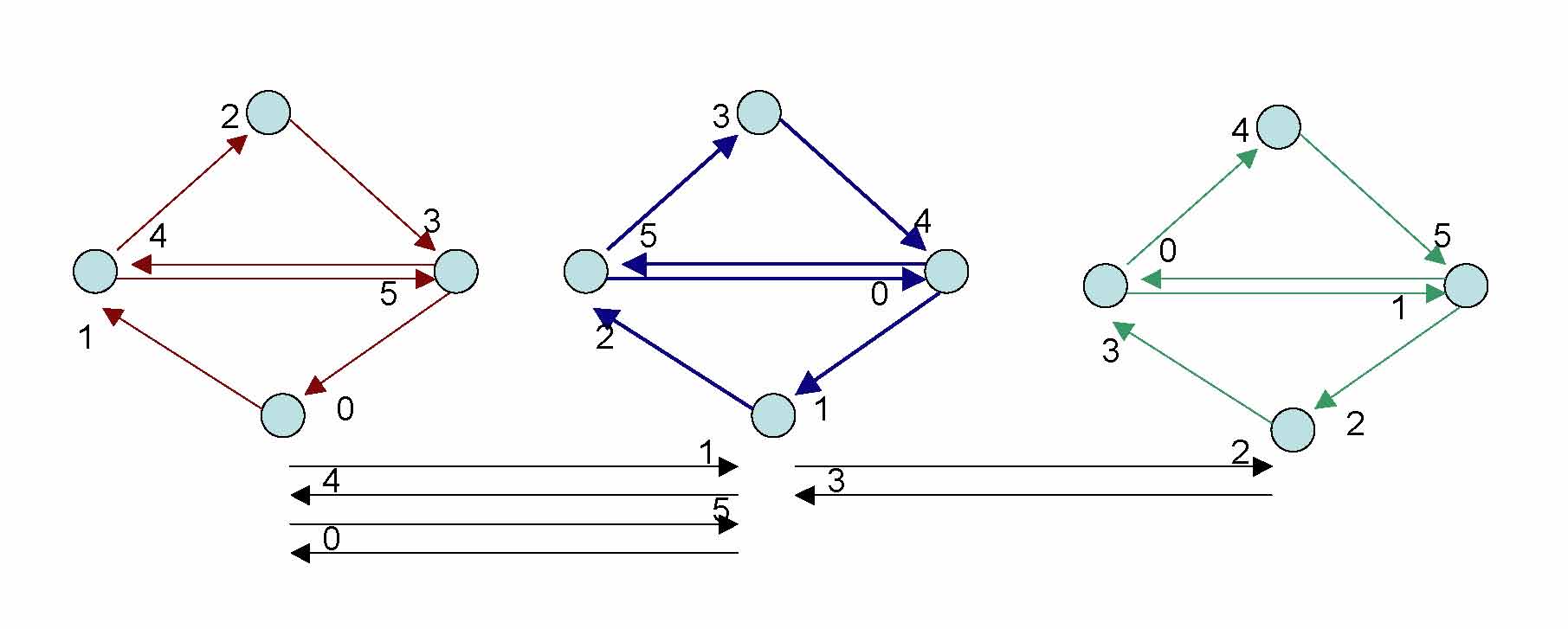}
 	\caption{\footnotesize PV graph of Theorem \ref{teo:moveLBarb} with  $n=12$, $k=4$, $p=6$.}\label{construction1}
 \end{figure}

  \begin{proof}  
  %{\bf (of Theorem \ref{teo:moveLBarb})}\\
%CONSTRUCTION 1  (''loop of loops") :
Let $S=\{s_0,\ldots,s_{n-1}\}$ and $C=\{c_0,...,c_{k-1}\}$.
Partition the set $S$ into $k-1$ subsets $S_0, \dots, S_{k-2}$ with
$|S_i| = \lfloor \frac{n}{k-1}\rfloor $ for $0\leq i \leq k-3$ and $S_{k-2}$
containing  the rest
of the elements. From each set $S_i$ select a site $x_i$;
let $X= \{x_0, \ldots, x_{k-2}\}$.
 For each $c_i$, $i<k-1$, construct a route $\pi(c_i)$
of period $p$ traversing $S_i$  and such that $x_i$ is visited only at time
$t \equiv i  \mod p$; this can always be done because   $|S_i|\geq 3$,
  since
  $k \leq \frac{n}{3}$.
   %$0\leq i \leq k-1$
Construct for $c_{k-1}$ a route $\pi(c_{k-1})$
of  period $p$ traversing $X$ such that  it visits $x_i$ at time
$t \equiv i  \mod p$ (it might visit it also at other times).
Thus, by construction,
carriers $c_i$ and $c_{k-1}$
have only one meeting point,  $x_i$, and 
%intersects $\pi(c_i)$ and $\pi(c_{k-1})
only at time $t \equiv i  \mod p$, 
while  $\pi(c_i)$ and $\pi(c_{j})$
have no meeting points at all,  $0\leq i \neq j \leq k-2$. See Figure \ref{construction1} for an example.
The agent \al must hitch a ride with every $c_i$ to visit the disjoint sets $S_i$,
$0\leq i\leq k-2$;
however,   \al can enter route $\pi(c_i)$ only at time $t \equiv i  \mod p$ and,
once it enters it, \al can leave it only after time $p$, that is only after the entire
route $\pi(c_i)$ has been traversed.
When traversing the  last set $S_i$, \al could stop as soon as all its
$|S_i| \geq   \lfloor \frac{n}{k-1}\rfloor $ elements are visited.
Additionally  \al must perform at least $k-2$ moves
on $\pi(c_{k-1})$ to reach each of the other routes.
In other words, \al must perform at least
$  (k-2) p\ +   \lfloor \frac{n}{k-1}\rfloor + (k-2)$ moves.

  \end{proof}
  \ \\ 
  
\noindent Costs can be significantly higher in heterogeneous systems as shown by
the following:

  \begin{theorem}
     \label{teo:moveLBarbHete}
        For any $n, k, p$, with $n\geq 9$, $
         \frac{n}{3}\geq k \geq 3$, and  $p \geq \max\{k-1,\lceil \frac{n}{k} \rceil \} $,
         there exists  a feasible {\em heterogeneous} graph $\vec{G}_R $ 
          with $n$ sites, $k$ carriers and period
        $p$ such that   
%        \begin{center}
  ${\cal M}(\vec{G}_R) \geq 
  (k-2) (p-1)p +\lfloor \frac{n-2}{k-1}\rfloor -1$.
%  \end{center}
    This result holds even if
%    the  system is {\em homogeneous},
     \al knows $ \vec{G}_R, k$ and $ p$, and has unlimited
    memory.
      \end{theorem}

\begin{proof} 
%{\bf of Theorem \ref{teo:moveLBarbHete})} \\
Let $C=\{c_0,...,c_{k-1}\}$.
Partition the set $S$ into $k$ subsets $S_0, \dots, S_{k-1}$ with
$|S_i| = \lfloor \frac{n-2}{k-1}\rfloor $ for $1\leq i \leq k-1$ and $S_{0}$
containing  the rest
of the elements.
From each set $S_i$   ($1 \leq  i\leq k-1$) select a site $x_i$;
let $X= \{x_1, \ldots, x_{k-1}\}$.
 For each $c_i$ ($1 \leq  i<k-1$), generate a route $\pi(c_i)$
of length $p$ traversing $S_i$  and such that $x_i$ is visited only at time
$t \equiv i  \mod p$; this can always be done because,
  since
  $k \leq \frac{n}{3}$,  we have  $|S_i|\geq 3$.
   %$0\leq i \leq k-1$
Construct for $c_{0}$ a route $\pi(c_{0})$
of  period $p-1$ traversing $S_0 \cup X$ such that  it visits $x_i\in X$
only at time $t \equiv i  \mod (p-1)$;  this can always be done
since $|S_0| + |X| \geq 2 + k- 1 = k + 1$.
In other words,
 in  the system  there is a route of period $p-1$,   $\pi(c_0)$,
and   $k-1$ routes of period $p$,  $\pi(c_i)$  for $0< i < k$.
Let \al be  at $x_0$ at time $t=0$;
it must hitch a ride with every $c_i$  ($0<i<k)$  to traverse the disjoint sets $S_i$;
 let $t_i$ denote the first time when \al hitches a ride
 with $c_i$.
 Since  $c_i$  has  connection only with $c_0$,
 to catch a ride on $c_i$ \al must be with $c_0$ when it meets $c_i$ at $x_i$
 at time $t_i$. To move  then to a different carrier
 $c_j$ ($i,j \neq 0)$,
 \al must
first return at $x_i$ and hitch a ride on $c_0$.
 Since $c_0$ is at $x_i$ only when  $t \equiv i  \mod (p-1) $
while  $c_i$  is there only when  $t \equiv i  \mod p$, and since
   $p-1$ and $p$ are coprime,  $c_0$  will meet  $c_i$    at time $t' > t_i$ if
and only if
 $t \equiv t_i  \mod(p\ (p-1))$.
In other words, to move from  $\pi(c_i)$ to another route
$\pi(c_j)$ \al must perform at least $p (p-1)$ moves.
Since \al must go on all routes,
at least $(k-2) p (p-1)$ moves must be performed until
\al hitches a ride on the last carrier, say $c_l$;
\al can stop only once
the last
unvisited sites in $\pi(c_l)$ have been visited, i.e., after
at least $\lfloor \frac{n-2}{k-1}\rfloor -1$ additional moves.
Therefore the number of moves \al must perform is at least
%\begin{center}
$(k-2) (p-1)p +\lfloor \frac{n-2}{k-1}\rfloor-1$,
% \end{center}
\noindent completing the proof.
\end{proof}

In other words,  by Theorems \ref{teo:moveLBarb} and \ref{teo:moveLBarbHete},
without any restriction on the routes, even if
 the  system is {\em homogeneous},
   \al knows $n, k, p$, and has unlimited
     memory
    \begin{equation}
{\cal M}_{homo}(n,k) =
  \Omega(k p) 
      \end{equation}
      \begin{equation}
    {\cal M}_{hetero}(n,k) = \Omega(k p^2)
    \end{equation}

Notice that the parameter $p$ in the above lowerbounds 
%  of
%  Theorems \ref{teo:moveLBarb}
%and \ref{teo:moveLBarbHete} 
can
be  arbitrarily large; in fact  a route can be arbitrarily long
even if its domain is small. 
This however can occur only if the carriers are allowed to go from a site $x$ to a site
$y$ an arbitrary amount of times  within the same
period.
Imposing restrictions on the amount of redundancy in the route the carriers must follow
will clearly have an
impact on the number of moves the agent needs to make.

 %------------------------------------------------------------
   \subsubsection{Simple Routes} \ \\
A natural restriction is that  each  route is
{\em simple}: the directed graph it describes  does not contain
self-loops nor multi-edges; that is, $\pi(c)[i]\neq\pi(c)[i+1]$  and,
if $\pi(c)[i]=\pi(c)[j]$ for $0 \leq i <j $, then $\pi(c)[i+1]\neq \pi(c)[j+1]$
where the operations on the indices are modulo $p(c)$.
If a route $\pi(c)$ is simple, then $p(c) \leq n (n-1)$.
Let us stress that even if all the routes are simple, the resulting
system $\vec{G}_R $ is not necessarily simple.

The routes used in the proof of Theorems \ref{teo:moveLBarb}
and \ref{teo:moveLBarbHete} were not simple.
The natural question is whether simplicity of the routes can lower the
cost fundamentally, i.e. to
$o(k p) \subseteq  o(k n^2)$ in case of homogeneous systems, and
to $o(k p^2) \subseteq  o(k n^4)$ in the heterogeneous ones. The answer is unfortunately negative
in both cases. \ \\

We will first consider the case of homogeneous systems with simple routes.

 \begin{theorem}
    \label{teo:moveLBsimp}
  For any  $n\geq 4$ and $ \frac{n}{2} \geq k \geq 2$
         there exists  a feasible {\em simple}  {\em homogeneous} PV-graph $\vec{G}_R $  with $n$ sites
          and $k$ carriers  such that  
%           \begin{center}
  ${\cal M}(\vec{G}_R) > \frac{1}{8} k n (n-8)$. 
%  = \Omega(n^2 k)$.
%  \end{center}
    This result holds even if
%     the  system is {\em homogeneous},
    \al knows $\vec{G}_R$ and $ k$, and has unlimited
     memory.
  \end{theorem}
  %----- IN APPENDIX

The proof can be found in the appendix.
Let us consider now the case of  heterogeneous systems with simple routes.

\begin{theorem}
    \label{teo:moveLBsimpHete}
  For any  $ n \geq 36$ and $\frac{n}{6}-2  \geq k \geq 4$ 
         there exists  a feasible {\em simple} {\em heterogeneous}
         PV-graph $\vec{G}_R $  with $n$ sites
          and $k$ carriers  such that   \begin{center}
  ${\cal M}(\vec{G}_R)  \geq \frac{1}{16} (k-3 ) (n^2 - 2n)^2$.
  \end{center}
    This result holds even if
   \al knows $\vec{G}_R$ and $ k$, and has unlimited
     memory.
  \end{theorem}
The proof can be found in the appendix.

 %------------------------------------------------------------
   \subsubsection{Circular Routes} \ \\ 
A  further restriction on a route is to be  {\em irredundant} (or {\em circular}): 
an edge appears in the route only once. In other words,
the resulting graph
is either a cycle
or a virtual cycle (i.e., induced by a simple traversal of a tree), hence
the name circular.

By definition, any circular route $\pi(c)$ is simple.
and $p(c) \leq 2(n-1)$. The system is irredundant if all the routes
are circular.
Let us stress that the fact that  the system is irredundant
does not imply that the
graph $\vec{G}_R $ is irredundant or even  simple.

        \begin{figure}[tbh]	
 	\centering
     \includegraphics[width=0.4\textwidth]{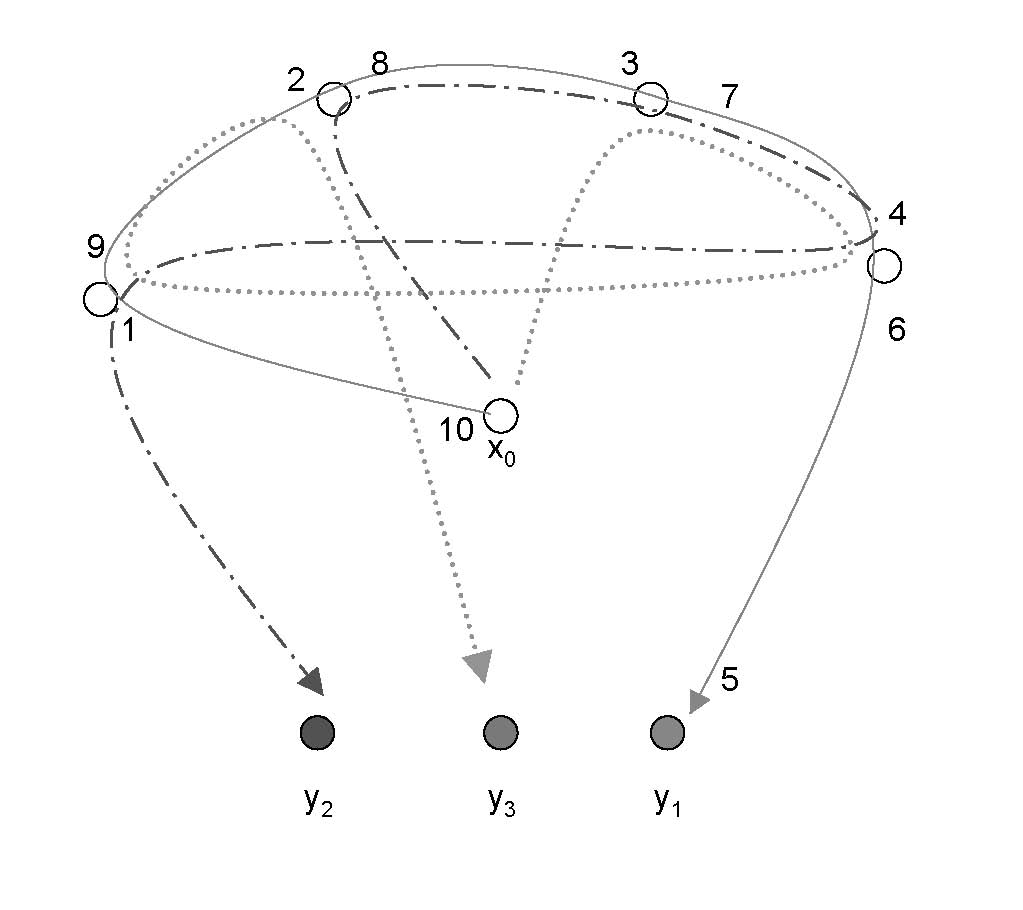}
 	\caption{\footnotesize $n=8$, $k=3$, $p=6$.}\label{construction2}
 \end{figure}
 
 The graph used in the proof of Theorem \ref{teo:moveLBsimp} is simple but not
irredundant. The natural question is whether irredundancy can lower the
cost fundamentally, i.e. to
$o(k p) \subseteq o(k n)$ for circular homogeneous systems and to
$o(k p^2) \subseteq o(k n^2)$ for circular  heterogeneous ones. 
The answer is unfortunately negative also in this case,
as shown in the following.

    \begin{theorem}
    \label{teo:moveLBirr}
     Let the systems  be {\em homogeneous}.
  For any  $n\geq 4$ and $ \frac{n}{2}\geq k \geq 2$
         there exists  a feasible {\em irredundant} simple graph $\vec{G}_R $  with $n$ sites
          and $k$ carriers  such that  
           \begin{center}
  ${\cal M}(\vec{G}_R) \geq n (k-1)$.
 \end{center}
    This result holds even if
     the  system is {\em homogeneous},
    \al knows  $\vec{G}_R$, $n$ and $ k$, and has unlimited
     memory.
  \end{theorem}
\begin{proof}
%CONSTRUCTION 2b: ("Homo Star")
Consider the system where $S= \{x_0, x_1, \ldots, x_{n-k-1}, y_1, y_2, \ldots, y_k\}$,
 $C= \{c_1, \ldots, c_k\}$, and the set of routes is defined as follows:

 $$\pi(c_i) =  \left\{\begin{array}{ll}
                     < x_0, \alpha(1),  y_{1}, \alpha(1)^{-1} >&\textrm{for $i=1$}\\
                      < x_0, \alpha(i), \beta(i),  y_{i }, \beta(i)^{-1}, \alpha(i)^{-1} > &\textrm{for $1< i \leq  k$}
                     \end{array}
                     \right.
$$

\noindent where  $\alpha(j) = x_{j}, x_{j+1}, x_{j+2}, \ldots, x_{n-k-1}$,
 $\beta(j) = x_{1},  x_{2}, \ldots,  x_{j-1}$, and $ \alpha(j)^{-1} $  and $\beta(j)^{-1}$
 denote the reverse of  $\alpha(j)$ and $ \beta(j)$, respectively.
In other words, the system is composed of   $k$ circular routes of period $p =2(n-k)$,
each with a distinguished site (the $y_j$'s);
the distinguished sites are
reached by the corresponding carriers simultaneously at time $t  \equiv n-k \mod p$.
The other $n-k-1$ sites are common to all routes; however
there is only a single meeting point in the system, $x_0$, and all
carriers reach it simultaneously at time $t \equiv 0 \mod p$.
More precisely, for all $1\leq i\neq j \leq k$, $c_i$ and $c_j$ meet only at $x_0$; this will
 happen whenever $t \equiv 0 \mod p$.\\
 Let \al start at $x_0$ at time $t=0$.
 To visit $y_i$, \al must hitch a ride on $c_i$; this can happen only at $x_0$ at time
 $t \equiv 0 \mod p$; in other words, until all $y_i$'s are visited, \al must
  traverse all $k$ routes (otherwise will not visit
all distinguished sites) returning to $x_0$; only once the last
distinguished site, say $y_j$ has been visited, \al can avoid returning to $a_0$.
Each route, except the last,  takes $2(n-k)$ moves; in the last, the agent can stop after
 only $n-k$ moves, for a total of   $2 k (n-k) - (n-k) $ moves. Since $k\leq \frac{n}{2}$,
$2 k (n-k) - (n-k) = 2nk - 2k^2 - n + k \geq (k-1)\  n $ and the Theorem follows.

\end{proof}

%{\bf QUESTION: can we make a construction when  $k> n/2$   ???}\\

We are now going to show that the cost can be
 order of magnitude larger  if
 the system is not homogeneous. The proof can be found in the appendix.

     \begin{theorem}
     \label{teo:moveLBirrHetero}
      Let the systems  be {\em heterogeneous}.
  For any  $0<\epsilon<1$, $ \frac{2}{\epsilon}\leq n$ and $ 2 \leq k \leq  \epsilon \ n$, 
         there exists  a feasible {\em irredundant}  graph $\vec{G}_R $  with $n$ sites
          and $k$ carriers  such that   
                    \begin{center}
   ${\cal M}(\vec{G}_R)  >   \frac{1}{4}\  (1-\epsilon)^2\ n^2\  (k-2)$
  $ =  \Omega(n^2 k)$.\\
  \end{center}
%    where  $\theta =  {(1-\epsilon)^2}/{4}$. 
    This result holds even if
    \al knows $\vec{G}_R$, $n$ and $ k$, and has unlimited
     memory.
  \end{theorem}

 %%%%%%%%%%%%%%%%%%%%%%%%%%%%%%%%%%%%%%
  \section{Optimal Explorations}
 %%%%%%%%%%%%%%%%%%%%%%%%%%%%%%%%%%%%%%

In this section we show that the limitations on computability and
complexity presented in the previous section are tight.
In fact we  prove that all necessary conditions are also sufficient
and all lower bounds on costs are tight.
We do so constructively presenting worst case optimal solution algorithms. 
An added benefit is that the algorithms
are rather simple.

We will first  introduce the notion
of {\em meeting graph}, 
that will be useful in the description and analysis of our exploration
algorithms. 
We will then describe and analyze  two  exploration
algorithms, one that does not require
unique node identifiers (i.e., the PV
graph could be {anonymous}), and one for the case when
distinct site ids are available.

%       \begin{figure}[tbh]	
% 	\centering
%     \includegraphics[width=0.8\textwidth]{meeting-graph2.jpg}
% 	\caption{\footnotesize (i)  A circular PV-Graph with  carriers $a$, $b$, $c$, $d$, and $e$;
%	$(ii)$ the corresponding meeting graph. The numbers represent  time.}\label{meeting}
% \end{figure}

The  {\em meeting graph} of  a PV graph  $\vec{G}$ is
the undirected graph     ${H(\vec{G})} = (C,E)$,
where each
 node corresponds to one of the $k$  carriers,  and there is an edge between 
 two nodes  if  there is at least 
 a meeting point between the two corresponding carriers.  
% As an example, consider the PV graph shown  in Figure \ref{meeting}. 
% In this example all the routes are simple cycles and they intersect in various points. 
%Note that, although routes 
%% BERNARD : shouldn't it be "d" instead of "c"
% $\pi(b)$ and $\pi(d)$ intersect, the two carriers never meet 
% because $b$ always passes by the possible intersection point 
% one time unit before $d$. On the other hand, although 
% $d$ and $e$  do  not meet within one cycle, they do connect
% every $p(d) \cdot p(e) = 3 \cdot 4 = 12$ time units.
%%   (where $p_D$ and $p_E$ 
%% are the periods respectively of $D$ and $E$).

%%---------------
%\subsection{Meeting Graph}
%%-----------------

%------------------------------------------------------------
 \subsection{Exploration of Anonymous PV Graphs}
\label{anonymous}

We first consider the general problem of exploring any
 feasible periodically varying graph without making any
 assumption on the distinguishability of the nodes.
By Theorem \ref{teo:anon-imp-p},  under these conditions the problem is not solvable
if an upper bound on the periods is not known to \al (even if \al has unbounded memory and knows $n$ and $k$).

We now prove that, if such a bound $B$  is known,
 any feasible periodically varying graph can be explored
even if the graph is anonymous, the system is heterogeneous, the routes  are arbitrary,
and  $n$ and $k$ are unknown to \alp
The proof is constructive: we present a simple 
and efficient exploration  algorithm
 for those conditions.

%The number of moves performed by the  algorithm is 
%worst case {\em optimal} whenever $B=O(p)$. Interestingly, 
%in this case, the move
%complexity remains worst-case optimal if, unknown to \al,
% the class of systems is restricted to simple PV graphs,
% or even to cyclic PV graphs.
%Furthermore, this is  done using only $O(\log B + k \log k)$ bits of memory.

Since the PV graph is anonymous and  $n$ and $k$ are  not known,
to ensure that no node is left unvisited, the algorithm will have
\al explore  all domains, according to a simple
but effective strategy; the bound
$B$  will be used to determine termination.
%:
%riding with a carrier for $B$ time units allows \al to fully visit 
%its domain. 

 Let us now describe the algorithm, {\sc Hitch-a-ride}.
The exploration strategy used by the algorithm is best described 
as a   {\em pre-order}  traversal of  
a {\em spanning-tree}  of the  meeting graph  $H$, where
"visiting"   a node of the 
meeting graph $H$  really consists  of riding   with 
the  carrier corresponding to that node for $B'$ time units,
where $B' = B$ if the  set of routes is known to be homogeneous,
$B'=B^2$ otherwise
(the reason for this amount will be apparent later).
%%The idea is for the agent to explore  the   domains  in such a way that
%%the corresponding meeting graph is  fully visited.
%In fact, a depth-first traversal of the meeting graph is performed by
% completing a full tour of  a domain when its carrier is taken for the first time, 
%%it is visited by following it for  $B$ time units  
%and then proceeding with a new encountered 
% carrier or backtracking  to one already visited if none is available.
 
 More precisely, assume that   agent \al  is    riding  with     $c$ for the first time;
 it will  do so for  $B'$ 
time units
 keeping   track of all new carriers  encountered
(list  $Encounters$). By that time,  \al has not only visited  the domain of $c$
but, as we will show, \al has encountered all carriers
that can meet with $c$ ( i.e., all the neighbours of $c$ in the
meeting graph $H$).

At this point  \al has "visited" $c$ in $H$; it will then continue the traversal of $H$
moving  to an unvisited neighbour; this is done by \al
continuing to ride with $c$ until a new carrier $c'$ is encountered;
$c$ will become the "parent" of $c'$.
If all neighbours of $c$ in $H$ have been visited, \al will return to
its "parent" in the traversal;  
this is done by \al
continuing the riding with $c$ until its parent is encountered.
 The algorithm terminates when \al returns to the starting carrier
 and the   list  $Encounters$ is empty.

\ \\
The formal recursive description of   Algorithm {\sc Hitch-a-ride}
%can be found in the Appendix. 
is given below.

%\noindent {\bf C1:  Algorithm {\sc Hitch-a-ride}} \\

Let \al start with carrier $c_0$.

%FIX THIS Note that this algorithm can be used when there are certain constraints on the type of domains.
%This is the case, for example,  of {\em simple} domains (where the bound is $n^2$)
% or domains without repetitions  of strings (where the bound is not a priori known but can be derived online).
Initially:    $Home = c_0$; $parent(Home):=\emptyset$
%$\forall c, parent-of-(c):=\emptyset$;  
$Visited := \emptyset$; $Encounters$  $:=\{c_0\}$; $N(c_0)=\emptyset$;\\

\begin{center} 
%Algorithm {\sc Hitch-a-ride}\\ \ \\
\fbox{
\begin{minipage}{5cm} 
  {\sc Hitch-a-ride}(c)  
 \begin{tabbing}
 {\bf  if}  $c= $  \= $Home$   and  $|Encounters| = \emptyset$ {\bf then}  \\
   \>  Terminate\\
{\bf else}\\
\>  {\bf  if}  \= $c \notin Visited$ {\bf then}  \\
\>\>   \sc{Visit}$(c)$\\
\>  {\bf  end-if} \\
\>    $c'  \leftarrow $  {\sc Go-to-Next}$(c)$\\
 \>  \sc{Hitch-a-ride}$(c')$
  \end{tabbing}
\end{minipage}
}
\end{center}

\begin{center}
\fbox{
\begin{minipage}{10cm}
  {\sc Visit}($c$)
 \begin{tabbing}
   $MyParent \leftarrow$ parent(c); 
   $N(c):=\{MyParent\}$\\
ride with $c$ \= for $B'$ time units, and while riding\\
  \> {\bf if} meet \=   carrier $ c' \notin (Encounters \cap Visited)$  {\bf then}\\
\> \> $Encounters:=\ Encounters\ \cup\ \{c'\}$  \\
\> \> $N(c):=N(c) \cup \{c'\}$  \\
\>  {\bf  end-if} \\
%----
%  {\bf if} \= $c \notin Visied$  {\bf then}\\
   $Visited :=\ Visited   \cup\ \{c\}$ \\
    $Encounters:=\ Encounters  - \ \{c\}$  
  \end{tabbing}
\end{minipage}
}
\end{center}

%---

\begin{center}
\fbox{
\begin{minipage}{10cm}
  {\sc Go-to-Next}$(c)$
 \begin{tabbing}
{\bf if} \=   $(N(c)\cap Encounters)  \neq \emptyset$ {\bf then}\\
\>   Continue the ride until meet  $c' \in (N(c) \cap Encounters) $ \\
\>  parent-of-(c'):= c\\
\> return $c'$ \\
  {\bf else} \=  \\
\>   Continue the ride until  encountering  $MyParent$ \\
\>   return $MyParent$
  \end{tabbing}
\end{minipage}
}
\end{center}

\begin{theorem}
\label{teo:anon-upper}
Algorithm {\sc Hitch-a-ride} correctly explores  any
feasible  PV graph in finite time provided (an
upper bound on) the size of largest route is known.
\end{theorem}

\begin{proof}  
%({\bf of Theorem \ref{teo:anon-upper})}\\
First observe that, when 
executing {\sc Visit}($c$), \al rides with $c$
for $B'$ time units, and by definition
$B' \geq B \geq p(c)$; thus, \al
would visit the entire domain of $c$.
Next observe that,
after the execution of  {\sc Visit}($c$), 
$N(c)$ contains the ids of all the
carriers that  have a meeting point with $c$.
In fact, any two routes  $\pi(c_i)$  and $\pi(c_j)$
 that have a common meeting point
will meet there every $p_{i,j}$ time units,
where $p_{i,j}$ is the least common multiple of
$p(c_i)$
and $p(c_j)$. 
If the set of routes is known to be homogeneous,
by definition  $\forall i,j$ 
$B'=B \geq   p_{i,j} = p(i) =p(j)$. 
If instead the set of routes is heterogeneous or
it is homogeneous but it is  not 
known to be so,  
by definition 
$\forall i,j$  $B'=B^2 \geq   p(i) \times p(j) \geq  p_{i,j}$. 
Hence 
by riding $B'$ time units with $c$, \al will encounter
all carriers  that  have a meeting point with $c$.
In other words, after the "visit" of a node in $H$,
\al knows all its neighbours, and which ones have not yet been
visited.
Thus,  \al will correctly perform
a pre-order visit of all the nodes of the spanning tree of $H$
rooted in $c_0$ defined by the relation "parent-of".
Since, as observed, the visit of a node in $H$ consists of
a visit of all the node in its domain, the theorem holds.
\end{proof}

\noindent This proves that the necessary condition for    {\em PVG-Exploration}
expressed 
by Theorem  \ref{teo:anon-imp-p}
is also sufficient.

Let us now consider the cost of the algorithm.

\begin{theorem}
\label{teo:anon-upb}
The number of moves performed by {\sc Hitch-a-ride} 
to traverse a feasible PV graph  $\vec{G}$  is 
at most  $(3 k - 2) B'$.
 where $k$ is the number of carriers and
$B'$ is the known (upperbound on the) size of the largest route.
\end{theorem}
\begin{proof}
Every time routine {\sc visit}$(c)$ is executed,  \al performs $B'$ moves;
since a visit is performed for each carrier, there will be a total of $k \cdot B'$ moves.
Routine {\sc Go-to-Next}$(c)$
is used to move from  a carrier $c$  to another $c'$
having a meeting point in common. This is achieved by riding with 
$c$ until $c'$ is met; hence its execution costs at most $B'$ moves.
The routine  is executed   to move from a carrier to each of
its "children", as well as to return to its
"parent" in the post-order traversal of the 
spanning tree of $H$ defined by the relation "parent-of". 
In other words, it will be executed precisely $2 (k-1)$ times
for a total cost of at most $ 2(k-1) B'$ moves.
The theorem then follows.
\end{proof}

The efficiency  of  Algorithm {\sc Hitch-a-ride}   clearly depends on the accuracy of the 
upperbound $B$ on the size  $p$  of the longest route in the system, as large values
of $B$ affect the number of moves  linearly in the case of homogeneous systems,
and quadratically in the case of heterogeneous system.
However, it is sufficient that the upperbound is linear in $p$ for the algorithm to
be {\em optimal}. In fact,
 from Theorem \ref{teo:anon-upb} and from the lowerbounds of
Theorems  \ref{teo:moveLBarb}-\ref{teo:moveLBirrHetero}
we have:

\begin{theorem}
\label{teo:anon-opt}
%\begin{enumerate} 
%\item Algorithm {\sc Hitch-a-ride} is optimal with respect to the amount
%of a priori information for the exploration of feasible
%anonymous PV graphs.  \ref{teo:anon-imp-p},
%\item 
Let $B=O(p)$; then Algorithm {\sc Hitch-a-ride} is worst-case optimal with respect to the amount
of moves. This optimality holds even if 
 (unknowingly)  restricted to  the class of feasible PV graphs {\em with ids}, 
 and even if the class is further restricted to be
{\em simple}  or
 {\em circular}  (anonymous or not).
%\begin{enumerate}
%\item unknowingly restricted to the class of feasible {\em non-anonymous} PV graphs
%\item unknowingly restricted to the class of feasible {\em simple} PV graphs
%\item unknowingly  restricted to the class of feasible {\em circular} PV graphs
%\end{enumerate}
%\end{enumerate}

\end{theorem}

It is interesting to note that
the amount of {\em memory} used by the algorithm
is relatively small: 
%which is optimal.
%\end{theorem}
%\begin{proof}
 $O(k \log k)$ bits are used to keep track of all the carriers and
 $O(\log B )$  bits to count up to $B^2$, for a
 total
 of  $O(\log B + k \log k)$ bits.
%\end{proof}

   %------------------------------------------------------------
   \subsection{Non-Anonymous Systems}
   %------------------------------------------------------------
 We now consider the case when the nodes have distinct Ids.
   By Theorem  \ref{teo:IdImpPn}, under these conditions,
   either $n$ or an upperbound on the system period must be available
   for the exploration to be possible.
   
   If an upperbound on the system period is available, the algorithm
   presented in the previous section would already solve the problem;
   furthermore, by Theorem \ref{teo:anon-opt}, it would do so optimally.
   Thus, we need to consider only the situation when 
   no upperbound 
   on the system period is available, and
   just $n$ is known.

 The exploration strategy  we propose
 %is similar to the one of the previous algorithm and 
 is based on a  {\em post-order}  traversal of  
a {\em spanning-tree}  of the  meeting graph  $H$,  where
"visiting"   a node $c$  of the 
meeting graph $H$  now  consists  of riding  
with $c$  for an amount  of time
large 
enough (1) to visit all the nodes in its domain, and (2) to meet every carrier
that has a meeting point in common with $c$. 
In the current setting, unlike the one considered previously,  an upper bound on the size of the 
domains is not available,  making the correct 
 termination of a visit problematic.
To overcome this problem, the agent will perform a sequence 
of {\em guesses} on the largest period $p$, each followed by a verification (i.e., a traversal).
If the  verification fails, a new (larger)  guess is made
and a new traversal    is  started.
The process continues  until $n$ nodes are visited,
 a detectable situation since nodes have ids.
%which could occur either when the guess
%is correct and thus the traversal in fully completed, or if the nodes happen
% to be visited with a wrong guess.    
%The agent \al takes new carriers in the order they are encountered
 
Let us describe the strategy more precisely.
Call a  guess  $g$ {\em ample}  if  $g\geq P$,
 where  $P=p$ if the graph is (known to be) homogeneous,  $P=p^2$ otherwise.
To explain how the process works, assume first that \al  starts the exploration riding 
with $c_0$ with  an ample guess $g$.
The algorithm would work as follows.
When   \al  is    riding  with  a carrier   $c$ for the first time,
 it will ride (keeping track of all visited nodes)    until   either it  encounters a new carrier $c'$ 
 or  it has made $g$ moves.
In the first case, $c$ becomes its  "parent"  and \al starts riding with $c'$.
In the latter, 
   \al  has ``visited" $c$, and will returns to its parent.
   Termination occurs when \al has visited $n$ distinct nodes.
With a reasonings similar to that used
for the   algorithm
of Section \ref{anonymous}, 
it is not difficult  to see that this strategy will allow \al to
correctly  explore the graph. 

Observe that this strategy might work even if 
 $g$ is not ample, since termination occurs once  \al detects that all $n$ nodes have been visited,
 and this might happen before all
nodes of $H$ have been visited. 
On the other hand, if  the (current) guess is not ample,
then the above exploration strategy might  not result in a full traversal,
and thus \al might not visit all the nodes.

Not knowing whether the current guess $g_i$  
is sufficient,  \al proceeds as follows: 
it  attempts to explore following the post-order traversal strategy
indicated above, but at the first indication that the guess is not
large enough, it   starts a new traversal using the current carrier
with a new guess $g_{i+1} > g_i$.
We have three situations when  the  guess is discovered to be not ample.
% (1) while riding with a carrier $c $ different from the starting carrier,
% $g_i$ time units have elapsed and during that time
%\al has not encountered the parent carrier $c'$ 
%(clearly $g_i$ units are not sufficient to complete the tour of the current domain); 
 (1) while returning to its parent,
 \al encounters a new carrier 
 (the route is longer than
$g_i$); (2) while returning to its parent,
more than $g_i$ time units elapse  (the route is longer than
$g_i$);  (3)
the traversal  terminates at the starting carrier, but the number 
of visited nodes is smaller than $n$.
 In these  cases the guess is doubled and a new traversal is started. 
 Whenever a new traversal is started, all variables are reset except
  for the set $Visited$ containing the already visited nodes.

The formal recursive description of   Algorithm {\sc Hitch-a-guessing-ride}
%can be found in the Appendix. 
is given below.

%\noindent {\bf C2:  Algorithm {\sc Hitch-a-guessing-ride}}  

Initially:    $Home = c_0$; $parent(Home):=Visited := \emptyset$
$Encountered$ $:=\{c_0\}$.

%Let \al start with carrier $c_0$.

%; $ParentEncountered(c_0):= false$.

\begin{center} 
%Algorithm {\sc Hitch-a-ride}\\ \ \\
\fbox{
\begin{minipage}{8cm} 
  {\sc Hitch-a-guessing-ride}(c)  
 \begin{tabbing}
 {\bf  if}  \=  $|Visited| = n$ {\bf then}  \\
 \>  Terminate\\
%   {\bf endif}\\
%---
 {\bf else} \= \\
%\>\> $ParentEncountered(c):= false$\\
\>  $c'  \leftarrow $  {\sc Go-to-Next}$(c)$\\
\> \sc{Hitch-a-guessing-ride}$(c')$
%   {\bf endif}
  \end{tabbing}
\end{minipage}
}
\end{center}

 %---
\begin{center}
\fbox{
\begin{minipage}{10cm}
  {\sc Go-to-Next}$(c)$  (* returns new carrier or parent *)
 \begin{tabbing}
 $MyParent \leftarrow$ parent(c); \\
{\bf ride}  \=with $c$  for $g_i$ time units, and while riding\\
\>  let $x$ be the current node, $Visited :=Visited \cup x$\\
%\>   {\bf if} meet \=   myParent:   \\
%\>\> $ParentEncountered(c):= true$\\ 
%\>   {\bf endif}\\
%----
\> {\bf if} meet \=   carrier $ c' \notin (Encountered )$  {\bf then}\\
\> \> $Encountered:=\ Encountered\ \cup\ \{c'\}$  \\
\>\> parent(c'):=c\\
\>\> Return(c')\\
%\>  {\bf  end-if} \\
%---
{\bf end-of-ride}  \\
 {\bf if}   ($c=Home$)  {\bf then} \\
\> {\bf if} ($|Visited| \neq n$) {\bf then}  \\
 \> \> {\sc Restart}$(c)$\\
 \>  {\bf else} \\
  \> \>  Terminate\\
 %{\bf if}  $ParentEncountered(c):= false$ {\bf then}\\
%  \> {\sc Restart}$(c)$\\
  {\bf else} \\
 \> $c' \leftarrow${\sc Backtrack}$(c)$\\
 \> Return(c')
% {\bf endif}\\
  \end{tabbing}
\end{minipage}
}
\end{center}

 %---

\begin{center}
\fbox{
\begin{minipage}{10cm}
  {\sc Backtrack}$(c)$  (* backtrack unless discover guess is wrong *)
 \begin{tabbing}
{\bf ride} \=  with $c$  until meet   $Myparent$ \\
\>  let $x$ be the current node, $Visited :=Visited \cup x$\\
  \> {\bf if}  wh\=ile  riding (encounter $c' \notin Encountered $) or ($g_i$ units elapse) \\
   \> \>  {\sc Restart}$(c)$ \\
 %\> {\bf endif}\\
 {\bf end-of-ride}\\
return $MyParent$
  \end{tabbing}
\end{minipage}
}
\end{center}
 
%--------------------------------------------------------

\begin{center}
\fbox{
\begin{minipage}{10cm}
{\sc Restart}$(c)$  (* reset variables except for $Visited$ *)
 \begin{tabbing}
 $guess:=2 \cdot guess$   (** new guess**) \\
 $Home := c$;  $parent(Home):=\emptyset$\\
$Encountered:=\{c\}$\\
  {\sc Hitch-a-guessing-ride}(c)  
  \end{tabbing}
\end{minipage}
}
\end{center}

\begin{theorem}\label{correct}
Algorithm {\sc Hitch-a-guessing-ride} 
correctly explores  any
feasible  PV graph with ids in finite time provided
the number of nodes is known.
\end{theorem}
 
\begin{proof} 
%({\bf of Theorem \ref{correct})} \\
Consider the case when  \al starts the algorithm from carrier $c_0$ with
an ample guess $g$.
  First observe that, when 
executing {\sc Go-to-next}($c$),   \al either
encounters  a new carrier and   hitches a ride with it, 
or it   traverses the entire domain of $c$ (because it rides with it
for $g \geq p(c)$ time units) before returning to its ``parent". 
Moreover, while traversing $c$, it does encounter all the carrier it can
 possibly meet. In fact, any two routes  $\pi(c_i)$  and $\pi(c_j)$
 that have a common meeting point,
will meet there every $p_{i,j}$ time units,
where $p_{i,j}$ is the least common multiple of
$p(c_i)$ and $p(c_j)$. 
If the set of routes is known to be homogeneous,
by definition  $\forall i,j$ 
$g \geq   p_{i,j} = p(i) =p(j)$. 
If instead the set of routes is heterogeneous or
it is homogeneous but it is  not 
known to be so,  
by definition 
$\forall i,j$  $g \geq   p(i) \times p(j) \geq  p_{i,j}$. 
Hence 
by riding $g$ time units with $c$, \al will encounter
all carriers  that  have a meeting point with $c$.
In other words, when    
executing {\sc Go-to-next}($c$),
 if \al does not  find new carriers  it "visits"  a node in $H$,
 and all its neighbours but its parent have been visited.
Thus,  \al will correctly perform
a post-order visit of all the nodes of a spanning tree of $H$
rooted in $c_0$.
Since, as observed, the visit of a node in $H$ consists in
a visit of all the node in its domain, the Lemma holds.

Let  the current guess $g_i$ be not ample.
This fact could be    detected
 by \al  because
while returning to the parent,
 \al encounters a new carrier or 
 $g_i$ time units elapse without encountering the parent.
 If this is the case, 
 \al will start a new traversal with the larger
 guess $g_{i+1}$.
Otherwise, \al  will returns to its starting carrier $c$  and 
complete its "visit"  of $c$. At this time, if all nodes have been visited,
\al will terminate (even if the guess is not ample); otherwise,
 a new traversal with the larger
 guess $g_{i+1}$ is started.
 That is, if  $g_i$ is not ample and there are still
 unvisited nodes, \al will start with a larger guess.
Since guesses are doubled at each restart,
after at most $\log P$ traversals, the guess will be ample.
\end{proof}

\noindent This theorem, together with 
Theorem \ref{teo:anon-upper}, proves that the necessary condition for    {\em PVG-Exploration}
expressed 
by Theorem  \ref{teo:IdImpPn}
is also sufficient.

Let us now consider the cost of the algorithm.

\begin{theorem}
\label{teo:id-upb}
The number of moves performed by
Algorithm {\sc Hitch-a-guessing-ride}  to traverse a feasible PV graph  $\vec{G}$  is $O(k\cdot P)$. 
\end{theorem}
\begin{proof}
First note that the worst case occurs when  the algorithm terminates with an ample
guess $g$. Let us consider such a case. 
%---
Let $g_0,g_1,\ldots, g_m=g$ be the sequence of guesses leading 
to  $g$  and
consider   the number of moves performed the first time \al uses an ample guess. 

 Every time routine {\sc Go-to-Next}$(c)$ is executed \al incurs in  at most $g_i$ 
 moves. Routine   {\sc Go-to-next}$(c)$    either returns a   new carrier 
  (at most $k$ times)  or  a "parent"  domain through routine  {\sc backtrack}(c) 
  (again at most $k$ times). Routine {\sc backtrack}(c)  spends at most $g_i$ moves every time it is called and it is called for each backtrack (at most $k$ times). So the overall move complexity 
  is    $ 3 g_i \cdot k$.
Let $g_0,g_1,\ldots, g_m$ be the sequence of guesses performed by the
algorithm.  Since the Algorithm correctly terminates if a guess is ample, only $g_m$
can be ample; that is    $ g_{m-1} < P  \leq g_m$.  Since $g_i = 2 g_{i-1}$, then
 the total number of moves will
be at most  $\sum_{i=0}^m  3 k g_i  < 6 k g_m = O(k \cdot P)$.
%

% Each preceding guess $g_i$ has generated at most  $O(g_i \cdot k)$ moves
% for a total of $\sum_{i=0^m} g_i k =  k\sum_{i=0^m} g_i = O(g_m \cdot k)=O(g \cdot k)$.
%Since, by definition, $g_{m-1} < P$, we have that 
%  $g \leq 2 P$ and thus  $O(g \cdot k) = O(P \cdot k)  $
\end{proof}

\begin{theorem}
\label{teo:id-opt}
%\begin{enumerate} 
%\item Algorithm {\sc Hitch-a-ride} is optimal with respect to the amount
%of a priori information for the exploration of feasible
%anonymous PV graphs.  \ref{teo:anon-imp-p},
%\item 
Let $B=O(p)$; then Algorithm {\sc Hitch-a-ride} is worst-case optimal with respect to the amount
of moves. This optimality holds even if 
 (unknowingly)  restricted to  the class of  {\em simple} feasible PV graphs {\em with ids}, 
 and even if the the graphs  in the class are further restricted to be
 {\em circular}.
%\begin{enumerate}
%\item unknowingly restricted to the class of feasible {\em non-anonymous} PV graphs
%\item unknowingly restricted to the class of feasible {\em simple} PV graphs
%\item unknowingly  restricted to the class of feasible {\em circular} PV graphs
%\end{enumerate}
%\end{enumerate}
\end{theorem}
%\begin{proof} It follows from Theorem \ref{teo:id-upb}  and from the lowerbounds of
%Theorems  \ref{teo:moveLBarb}-\ref{teo:moveLBirrHetero}.
%\end{proof}

\noindent The proof follows from Theorem \ref{teo:id-upb}  and from the lowerbounds of
Theorems  \ref{teo:moveLBarb}-\ref{teo:moveLBirrHetero}.
  
Finally, notice that 
the amount of {\em memory} used  by the algorithm is rather small: 
 $O(n \log n)$ bits
to keep track of all the visited nodes.

%------------------------------------------------------------
%   \section{Concluding remarks}
   %------------------------------------------------------------

\noindent {\bf Acknowledgments.}
We would like to thank David Ilcinkas for the helpful comments.

%%%%%%%%%%%%%%%%%%%%%%%%%%%%%%%%%%BIBLIO

\newpage
  \pagenumbering{roman}

\begin{center}
{\Large \bf  APPENDIX}
\end{center}

%\ \\

%\ \\
%\noindent {\bf A:  TABLES - SUMMARY OF RESULTS} \\

%\newpage

\noindent  {\bf B: PROOFS OF THEOREMS AND LEMMAS} \\

 \begin{proof}    {\bf (of Theorem  \ref{teo:moveLBsimp})} \\
To prove  this theorem we will first construct a system
satisfying the theorem's hypothesis.
Let
 $C= \{c_1, \ldots, c_k\}$,
 $S= \{x_0, \ldots, x_{\bar{m}-1}, y_1, y_2, \ldots, y_k, z_1, ..., z_{\bar{n}}\}$,
 where 
 $\bar{m} = \max\{i < n-k$: $i$  is prime$\}$, 
 and let $ \bar{n}=  n  - \bar{m}-k$.
 Consider the set of indices  $\iota(i,j)$
% $(1\leq i \leq k,  1\leq j \leq   \bar{m}^2 -  \bar{m} +1 )$,
defined as follows, where all operations
  are modulo  $\bar{m}$: for $0\leq s\leq\bar{m}-2$, $ 0\leq r\leq \bar{m}-1$ and $1\leq i\leq k$
 \begin{equation}
 \label{eq:iota}
% $\iota(i, 1) = \iota(i, \bar{m} l ) = i$   $(1\leq l \leq \bar{m}-1)$ \\
\iota(i, \bar{m} s + r ) =    i + (s+1) r
\end{equation}
For simplicity, in the following we will denote $x_{\iota(i,j)}$ simply as  $x(i,j)$.
Finally, let
the set of routes be defined as follows:
 \begin{equation}
  \label{eq:route}
  \pi(c_i)= < \mu, \delta(i), y_i>
 \end{equation}
 where
 \begin{equation}
  \label{eq:mu}
 \mu = z_1, ..., z_{\bar{n}}
\end{equation}
 and
 \begin{equation}
  \label{eq:delta}
 \delta(i) = x(i,1), x(i,2), \ldots, x(i,   \bar{m}^2 -  \bar{m} ).
 \end{equation}

The system {\tt SiHo} so defined is clearly homogeneous,

  \begin{claim}
  \label{lm:SimpleHomo1}
In {\tt SiHo}, for $1\leq i\leq k$,  $\pi(c_i)$ is {\em simple} and
 $p(c_i) = p =  \bar{m}^2 -  \bar{m} + 1 + \bar{n}$.
  \end{claim}
\begin{proof}
 That the value of $p(c_i)$ is as stated
 follows by construction.
  To prove simplicity  we must show that each edge in the route
 appears only once; that is, for all $1\leq i\leq k$,   $0 \leq t'< t''\leq p-1$,
  if $\pi(c_i)[t'] =  \pi(c_i)[t'']$ then  $\pi(c_i)[t'+1] \neq  \pi(c_i)[t''+1] $.
  This is true by construction for $t'<  \bar{n}$ and $t''\geq p-1$; i.e.,
  for the edges $ (z_1, z_2),(z_2, z_3), ..., (z_{ \bar{n}}, z_1), (z_1, x(1,1)), (x(i,p-2), y_i),
  (y_i, z_1)$.  Consider now the other values of $t'$ and $t''$.
Let $  \bar{n}  \leq t' = \bar{m} s' + r'  < \bar{m} s'' + r'' = t'' \leq p-2$ with
$\pi(c_i)[t'] =  \pi(c_i)[t'']$; that is
 \begin{equation}
 \label{eq:imp1}
   i + (s' +1) r' \equiv  i + (s'' +1) r''    \mod  \bar{m}
 \end{equation}
By contradiction,
 let   $\pi(c_i)[t'+1] =  \pi(c_i)[t''+1] $;
 that is
 \begin{equation}
 \label{eq:imp2}
    i + (s' +1) (r' +1) \equiv i + (s''  +1) (r'' +1)  \mod  \bar{m}.
 \end{equation}
But (\ref{eq:imp1}) and (\ref{eq:imp2}) together imply that  $s' \equiv s''$ (mod $ \bar{m}$), which in
turn (by  (\ref{eq:imp1})) implies that
 $r' \equiv r''$  (mod $ \bar{m}$).  However,
since $ \bar{m} $ is prime,  this can occur only if $s'=s''$ and $r'=r''$,
i.e. when  $t'=t''$;  a
contradiction.
\end{proof}

 \begin{claim}
 \label{lm:SimpleHomo2}
In {\tt SiHo},  $\forall i,j$ $(1\leq i < j \leq k)$, $c_i$ {\em and} $c_j$
 meet only at the nodes of $\mu$; will
 happen whenever $t \equiv l$\  (mod$\  p)$,  $0\leq l \leq  \bar{n}-1$
  \end{claim}
\begin{proof}
By definition, the carriers meet  at the nodes of $\mu$ only at the time stated by the lemma:
$\mu$ is the first part of each route, and the sites in $\mu$  are different from all the others.
To complete the proof we must show that two carriers will never meet anywhere else.
Since $y_i$ is only in route $\pi(c_i)$,  carriers never meet there. Let us
consider now the $x_i$'s.  By contradiction, let  $ \pi(c_i)[t] = \pi(c_l)[t]$
 for  some $i, l, t$ where $1\leq i\neq l \leq k$, $     \bar{n} \leq  t \leq p-1$; in other words,
 let
 $x(i,t) =   x(l,t)$.
The function $\iota$, by definition, is such that
$\iota(i+1,t) = \iota(i,t)+ 1 \mod \bar{m}$; since $\bar{m}$ is prime, this means that
$\iota(i,j) \neq  \iota(l,j) \mod \bar{m}$ for $1\leq i < l \leq k$ and $1\leq j \leq p-1$.
Therefore $\iota(i,t) \neq  \iota(l,t) \mod \bar{m}$; that is
$x(i,t) \neq  x(l,t)$:  a contradiction.
\end{proof}
\ \\

%We can now proceed with the proof of Theorem \ref{teo:moveLBsimp}:

%   \begin{theorem}
%    \label{teo:moveLBsimp}
%     Let the systems  be {\em homogeneous}.
%  For any  $n$ and $k \leq \frac{n}{2}$
%         there exists  a feasible {\em simple} graph $\vec{G}_R $  with $n$ sites
%          and $k$ carriers  such that   \begin{center}
%  ${\cal M}(\vec{G}_R) > \frac{1}{8} k n (n-8)$
%%  = \Omega(n^2 k)$.
%  \end{center}
%    This result holds even if
%%     the  system is {\em homogeneous},
%    \al knows $n$ and $ k$, and has unlimited
%     memory.
%  \end{theorem}

%  \begin{proof} (of Theorem \ref{teo:moveLBsimp}) \\
By Claims \ref{lm:SimpleHomo1} and
\ref{lm:SimpleHomo2}, the  {\tt SiHo}  system is composed of   $k\geq 2$ simple routes of period
 $p=  \bar{m}^2 -  \bar{m} + 1  -\bar{n} $,
each with a distinguished site (the $y_j$'s).
The other $n-k$ sites are common to all routes; however
the only meeting points in the system are those in $\mu$ and
each of them is  reached  by all
carriers simultaneously.
Let \al start at $z_1$ at time $t=0$. Since
 only $c_i$ can reach $y_i$,
 to visit all the distinguished sites $y_1, y_2, \ldots, y_k$, \al must hitch a ride on all carriers.
However, 
by Lemma \ref{lm:SimpleHomo2} carriers only connect  at the points of $\mu$,
 each of them  reached by  all carriers simultaneously.
 Thus, to visit $y_i$, \al must hitch a ride on $c_i$  at  a site in $\mu$ at time
 $t \equiv f \mod p$ for some $f\in\{0,..., \bar{n}-1 \}$. After the visit,
% In other words, until all $y_i$'s are visited,
  \al must
 return to $z_1$,  traverse all of $\mu$ hitching a ride on another carrier and
 follow that route until the end; only once the last
distinguished site has been visited, \al could stop, without returning to $z_1$.
In other words, to visit each $y_i$ (but the last),  \al will perform
 $p$ moves; in the visit of the last
distinguished site \al could stop after
 only $p- \bar{n} $ moves; in other words, \al
 needs to perform at least 
 $ (k-1) p + p -  \bar{n}  = k p -   \bar{n} $ moves. From Lemma~\ref{lm:SimpleHomo1},
% Since $p=  \bar{m}^2 -  \bar{m} + 2 + n - \bar{n}$
  it follows that

  \begin{center}

  $k p - (  \bar{n}) = k\  ( \bar{m}^2 -  \bar{m} + 1 +   \bar{n})   -    \bar{n} 
  >  k\  ( \bar{m}^2 -  \bar{m}) $
%  = k\  ( (\bar{n}-k)^2 -  (\bar{n}-k))$

% $  k p - 1 = k  (n-k)^2 + 2 k - 1 =  k n^2 - 2 n k^2 + k^3 +2 k^2 + k - 1 >
% k n ( n - 2 k) \geq  \frac{1}{4} k n^2 $.
 \end{center}

 \noindent Observe that, by definition of $\bar{m} $,  we have  $\frac{1}{2} (n-k-1) \leq \bar{m} \leq n-k-1$;
 furthermore, by hypothesis
  $ k \leq \frac{n}{2}$. Thus

   \begin{center}

  $ k\  ( \bar{m}^2 -  \bar{m})  \geq  k\ (\frac{1}{4} (n-k-1)^2 -  \frac{1}{2} (n-k-1))  =
 \frac{1}{4} k  (n-k)^2  - k (n-k) + \frac{3}{4} k $\\
 $>    \frac{1}{4} k  (n-k)^2  - k n
 \geq \frac{1}{8} n^2 k - k n $
% = \frac{1}{8} k n (n-8)$

  \end{center}

\noindent and the theorem holds.

\end{proof}
\ \\ 

%%%%%%%%%%%%%%%%%%%%%%%
\begin{proof} {\bf (of Theorem \ref{teo:moveLBsimpHete})} \\
To prove  this theorem we will first construct a system
satisfying the theorem's hypothesis.
Let
 $C= \{c_0, \ldots, c_{k-1}\}$,
% BERNARD surely it has to be max  $\bar{m} = \min\{q \leq  \frac{1}{2}(n-3k-4) :  q$ is prime$\}$ be the largest prime number smaller than
  $\bar{m} = \max\{q \leq  \frac{1}{2}(n-3k-4) :  q$ is prime$\}$,
%  be the largest prime number smaller than
%    $\frac{1}{2}(n-3k-4)$,
  and let $\bar{n} = n - 3k - 4 - 2 \bar{m}$. Observe that, by definition,
  \begin{equation}
  \bar{m} \geq \lceil \frac{\bar{n}}{2} \rceil
  \end{equation}
  Partition $S$ into six sets: $U = \{u_1, ..., u_{k-1}\}$, $V= \{v_1, \ldots, v_{k-2}\}$,
$W= \{ w_1, ..., w_{\bar{n}}\}$,  $X= \{x_1, \ldots, x_{\bar{m}}\}$, $Y=  \{ y_1, \ldots, y_{\bar{m}}\}$, and
 $Z = \{ z_1, ..., z_{k-1}\}$.
% $(1\leq i \leq k,  1\leq j \leq   \bar{m}^2 -  \bar{m} +1 )$,
Let the set of indices  $\iota(i,j)$ be as defined in  (\ref{eq:iota});
%for the As  homogeneous system  {\tt SiHo};
% $(1\leq i \leq k,  1\leq j \leq   \bar{m}^2 -  \bar{m} +1 )$,
%defined as follows, where all operations
%  are modulo  $\bar{m}$: 
% 
%  for $0\leq s\leq\bar{m}-2$, $ 0\leq r\leq \bar{m}-1$ and $1\leq i\leq k$
% \begin{equation}
% \label{eq:iota}
%% $\iota(i, 1) = \iota(i, \bar{m} l ) = i$   $(1\leq l \leq \bar{m}-1)$ \\
%\iota(i, \bar{m} s + r ) =    i + (s+1) r
%\end{equation}
for simplicity, in the following we will denote $x_{\iota(i,j)}$ 
and $y_{\iota(i,j)}$ simply as  $x(i,j)$ and $y(i,j)$, respectively.

%

%defined as follows, where all operations
%  are modulo  $\bar{m}$: for $0\leq s\leq\bar{m}-2$, $ 0\leq r\leq \bar{m}-1$ and $1\leq i\leq k$
% \begin{equation}
% \label{eq:psi}
% \psi(\bar{m} s + r ) =   (s+1) r
% \end{equation}
%In the following, for simplicity,  we will denote $x_{i+\psi(j)}$ and    $y_{i+\psi(j)}$
%simply as  $x(i,j)$ and $y(i,j)$, respectively.

%In the following, for simplicity,  we will denote $x_{i+\psi(j)}$ and    $y_{i+\psi(j)}$
%simply as  $x(i,j) =  x_{i+ } $ and $y(i,j)$, respectively.

%  We will consider two cases depending on the parity of  $n - k $.
%
%{\em Case 1.}  Let  $n-k$ be even.

Let the routes $R = \{\pi(c_0), ..., \pi(c_{k-1}) \}$ be defined as follows:

 \begin{equation}
  \label{eq:RoutesSimpleHetero}
 \pi(c_i)  =  <  \alpha(i), \gamma(i), \delta(i), \zeta(i) > 
\end{equation}

\noindent  where

  \begin{center}
%  \label{eq:routesimple}
$$  \alpha(i) = 
 \left\{\begin{array}{ll}
                        x(0,1), x(0,2), \ldots,   x(0, \  \bar{m}^2-\bar{m}-\lceil  \frac{\bar{n}}{2} \rceil )   &\ \textrm{for $ i=0$}\\
                 y(i,1), y(i,2), \ldots,   y(i, \  \bar{m}^2-\bar{m}- \lfloor  \frac{\bar{n}}{2} \rfloor-i+1)
                  &\ \textrm{for $ 0<i<k $}
                     \end{array}
                     \right.
$$
 \end{center}

%  \begin{center}
%%  \label{eq:routesimple}
%$$  \alpha(i) = 
% \left\{\begin{array}{ll}
%                        x(0,1), x(0,2), \ldots,   x(0,  \bar{m} ( \bar{m} -2) + \bar{m} - \lceil  \frac{\bar{n}}{2} \rceil )   &\textrm{for $ i=0$}\\
%                 y(i,1), y(i,2), \ldots,   y(i,  \bar{m} ( \bar{m} -2) + \bar{m} - \lfloor  \frac{\bar{n}}{2} \rfloor - i+1)
%                  &\textrm{for $ 0<i<k $}
%                     \end{array}
%                     \right.
%$$
% \end{center}

    \begin{center}
%  \label{eq:routesimple}
$$  \gamma(i) = 
  \left\{\begin{array}{ll}
w_1, w_2, \ldots, w_{  \lceil  \frac{\bar{n}}{2} \rceil } &\  \textrm{for $ i=0$}\\
                      w_{  \lceil  \frac{\bar{n}}{2} \rceil +1 }, w_{  \lceil  \frac{\bar{n}}{2} \rceil +2 }, ...,
 w_{ \bar{n} } &\ \textrm{for $ 0<i<k $}
                     \end{array}
                     \right.
$$
 \end{center}

   \begin{center}
%  \label{eq:routesimple}
$$  \delta(i) =   
\left\{\begin{array}{ll}
\emptyset  &\  \textrm{for $ i \leq 1$}\\
%                      y(i,  \bar{m} ( \bar{m} -1) - (i-1)), \ldots,  y(i,  \bar{m} ( \bar{m} -1))
 y(i,\  \bar{m}^2-\bar{m}-\lfloor  \frac{\bar{n}}{2} \rfloor -i+2), \ldots,  y(i,  \bar{m}^2-\bar{m})
                      &\ \textrm{for $ 1<i<k $}
                     \end{array}
                     \right.
$$
 \end{center}

%   \begin{center}
%%  \label{eq:routesimple}
%$$  \delta(i) =   
%\left\{\begin{array}{ll}
%\emptyset  & \textrm{for $ i \leq 1$}\\
%%                      y(i,  \bar{m} ( \bar{m} -1) - (i-1)), \ldots,  y(i,  \bar{m} ( \bar{m} -1))
% y(i,  \bar{m} ( \bar{m} -2) + \bar{m} - \lfloor  \frac{\bar{n}}{2} \rfloor -  i+2), \ldots,  y(i,  \bar{m} ( \bar{m} -1))
%                      &\textrm{for $ 1<i<k $}
%                     \end{array}
%                     \right.
%$$
% \end{center}

   \begin{center}
%  \label{eq:routesimple}
$$  \zeta(i) =   \left\{\begin{array}{ll}
z_1, z_2, \ldots,  z_{k-1}  &\ \textrm{for $ i = 0$}\\
u_1, z_1, v_1, \ldots, v_{k-2}  &\ \textrm{for $ i = 1$}\\
u_i, v_{k-2-i+2}, \ldots, v_{k-2},  z_i, v_1,\ldots, v_{k-2-i+1} &\ \textrm{for $ 1<i<k-1 $}\\
u_{k-1}, v_1, \ldots, v_{k-2}, z_{k-1}   &\ \textrm{for $ i = k-1$}\\
                     \end{array}
                     \right.
$$
 \end{center}

% where $[ .  ]^j$ denotes the $j$-th cyclic shift of the argument,
\noindent  and all operations on the indices are modulo $ \bar{m}$.
% For example, let $k=3, n=30$; then $|Y|=\bar{m}=5$ and $\bar{n}= 7$.
% For simplicity, let us denote $s_j\in S$ simply as $j$ $(0\leq j < 30)$,
% and consider the partition of $S$ as follows: $Y=\{1,2,3,4,5\}$, 
% 
% \noindent
% $\alpha(1)\gamma(1) = y_1,y_2,y_3,y_4,y_5,y_2,y_4,y_1,y_3,y_5,y_3,y_1,y_4,y_2,y_5,y_4,y_3,y_2,
% w_5,w_6,w_7,y_1,y_5$\\
%  $\alpha(2)\gamma(2) = y_2,y_3,y_4,y_5,y_1,y_3,y_5,y_2,y_4,y_1,y_4,y_2,y_5,y_3,y_1,y_5,y_4,
% w_5,w_6,w_7,y_3,y_2,y_1$\\
%   $\alpha(3)\gamma(3) =y_3,y_4,y_5,y_1, y_2,y_4,y_1,y_3,y_5,y_2,y_5,y_3,y_1,y_4,y_2,y_1,
% w_5,w_6,w_7,y_5,y_4,y_3,y_2$\\
% $\alpha(1)\gamma(1) = y_1,y_2,y_3,y_4,y_5,y_2,y_4,y_1,y_3,y_5,y_3,y_1,y_4,y_2,y_5,y_4,y_3,y_2,
% w_5,w_6,w_7,y_1,y_5$\\
%  $\alpha(2)\gamma(2) = y_2,y_3,y_4,y_5,y_1,y_3,y_5,y_2,y_4,y_1,y_4,y_2,y_5,y_3,y_1,y_5,y_4,
% w_5,w_6,w_7,y_3,y_2,y_1$\\
%   $\alpha(3)\gamma(3) =y_3,y_4,y_5,y_1, y_2,y_4,y_1,y_3,y_5,y_2,y_5,y_3,y_1,y_4,y_2,y_1,
% w_5,w_6,w_7,y_5,y_4,y_3,y_2$\\
% while \\
% $\delta(1) = 
The system {\tt SiHe} so defined has  the following properties:

   \begin{claim}
  \label{lm:SimpleHetero1}  In  {\tt SiHe}, for $0 \leq i \leq k-1$,  $\pi(c_i)$ is {\em simple}, and
$$ p(c_i)  =  \left\{\begin{array}{ll}
                       \bar{m}^2 - \bar{m} + k - 1 &\ \textrm{if $ i=0$}\\
                        \bar{m}^2 - \bar{m} + k  &\ \textrm{if $ 0<i<k $}
                     \end{array}
                     \right.
$$
  \end{claim}
\begin{proof}
 That the value of $p(c_i)$ is as stated
 follows by construction.
%  (i.e., from Expressions (\ref{eq:alpha})-(\ref{eq:delta})).
  To prove simplicity  of $p(c_i) $  we must show
 that, for all $0 \leq i\leq k-1$ and   $0 \leq t'< t''\leq p(c_i)-1$,
  if $\pi(c_i)[t'] =  \pi(c_i)[t'']$ then $\pi(c_i)[t'+1] \neq  \pi(c_i)[t''+1] $.

This is  true if one or more of
  $\pi(c_i)[t'] $,  $\pi(c_i)[t'+1] $, $ \pi(c_i)[t'']$,  $\pi(c_i)[t''+1] $  are in $\gamma(i)$ or $\zeta(i)$.
  In fact, by definition,  all the sites of $\gamma(i)$ and $\zeta(i)$ 
  ($Z$, half the elements of $W$, and if $i>0$ also $u_i\in U$ ) 
%  appear  in  $\gamma(i)$ and $\zeta(i)$ without any repetition,
%  and  are  neither in $\alpha(i)$ nor in
%  $\delta(i)$; that is, they
   appear in $\pi(c_i)$  without any repetition, i.e.,
   only once.

  Consider now all the other cases.
Let  $i, t', t''$ ( $0 \leq i\leq k-1$ and   $0 \leq t'< t'' < p(c_i)-2$) be
such that
 $\pi(c_i)[t'] =  \pi(c_i)[t'']$ but
 none of $\pi(c_i)[t'] $,  $\pi(c_i)[t'+1] $, $ \pi(c_i)[t'']$,  $\pi(c_i)[t''+1] $ are in $\gamma(i)$
 or in $\zeta(i)$.
  Let $t'= \bar{m} s' + r' $ and $t''= \bar{m} s'' + r''$.

Let
 $i>0$ (respectively, $i=0$);
 that is,
  $\pi(c_i)[t'] = y(i,t') = 
   y_{\iota(i,t')} = 
   y_{i+ (s'+1) r'} $ and $  \pi(c_i)[t'']=y(i,t'') =
 y_{\iota(i,t'')}  = 
 y_{i+ (s''+1) r''} $
 (respectively,
 $\pi(c_i)[t'] = x(0,t') = 
  x_{\iota(0,t')} = 
  x_{(s'+1) r'} $ and $  \pi(c_i)[t'']=x(0,t'') =
 x_{\iota(0,t''')}  =
  x_{(s''+1) r''} $).
 Since
 $\pi(c_i)[t'] =  \pi(c_i)[t'']$ it follows that
 $  y_{i+ (s'+1) r'} = y_{i+ (s''+1) r''} $
  (respectively,
$  x_{ (s'+1) r'} = x_{(s''+1) r''} $); that is,
 \begin{equation}
  \label{eq:SimpleHeteroImp1}
   (s' +1) r' \equiv   (s'' +1) r''    \mod  \bar{m}
 \end{equation}
 By contradiction,
let
 $\pi(c_i)[t'+1] =  \pi(c_i)[t''+1] $;
%Consider first the case  when $t' < t'' <  \bar{m} ( \bar{m} -2) +1$,
%and l
then
 \begin{equation}
  \label{eq:SimpleHeteroImp2}
   (s'+1) (r' + 1) \equiv   (s'' +1) (r'' + 1)  \mod  \bar{m}
 \end{equation}
But (\ref{eq:SimpleHeteroImp1}) and (\ref{eq:SimpleHeteroImp2}) together imply that  $s' \equiv s"$ (mod $ \bar{m}$), which in
turn implies that
 $r' \equiv r''$  (mod $ \bar{m}$).  However,
since $ \bar{m}$ is prime,  this can occur only if $s'=s''$ and $r'=r''$,
i.e. when  $t'=t''$;  a
contradiction.
%
% Consider now the case when $i=0$; once $y$ is substituted with $x$,
% exactly the same argument applies, yielding  also in  this case
% a proof by contradiction.
\end{proof}

    \begin{claim}
  \label{lm:SimpleHetero2}
In {\tt SiHe},  $\forall i,j$ $(1\leq i < j \leq k)$,
\begin{enumerate}
\item
 $c_i$ can meet with $c_0$  only at  $z_i$,
 \item
 $c_i$ {\em and} $c_j$ never meet.
 \end{enumerate}
%  this will
% happen at time $t$ if and only if  $t \equiv $\  (mod$\  p)$,  $0\leq j\leq n-\bar{n}+1$
  \end{claim}
\begin{proof}
First observe that (1) follows by construction, since $z_i$ is the only site in common between
$\pi(c_0)$ and $\pi(c_i)$, $i>0$.
To complete the proof we must show that any other two carriers,  $c_i$   and $c_j$
$(1\leq i < j \leq k)$,  will never meet; that is, $\pi(c_i)[t]\neq \pi(c_i)[t]$ for all $0\leq t \leq p-1$, where
$p = p(c_i) = p(c_j) =   \bar{m} ( \bar{m}-1)  + k $ (by Lemma \ref{lm:SimpleHetero1}).

By contradiction, let $\pi(c_i)[t]=\pi(c_j)[t]= s\in U\cup V \cup Y \cup Z \cup W$ 
for some $t <   p$.  \\
First observe that,  by construction,    $c_i$ visits only a single distinct element  
of $U$, $u_i\neq u_j$, and only a  single site in $Z$, $z_i\neq z_j$. Thus, $s\notin U\cup Z$.\\
Assume  $s= v_l\in V$.   By construction,   $ \pi(c_i)[t] =  v_l$ means that  
$  t = \bar{m}  ( \bar{m} -1)  +(  ( i + l ) \mod(k-1))$; on the other hand, $ \pi(c_j)[t] =  v_l$ means
 by construction that 
$  t = \bar{m}  ( \bar{m} -1)  +(  ( j+ l ) \mod(k-1))$.  Thus $( i + l ) \equiv ( j + l ) \mod(k-1))$
implying $i \equiv j \mod (k-1)$; but since $i<j\leq k-1$ it follows that
$i=j$, a contradiction. Hence $s\notin V$.\\
Assume now $s\in Y$.  Let  $t  =  \bar{m} l + r$.
By definition, $ \pi(c_i)[t]= \pi(c_j)[t]\in Y$ means that
  $y_{\iota(i,t)} =  y(i,t) =  \pi(c_i)[t] = 
  \pi(c_j)[t] =  y(j,t) = 
  y_{\iota(j,t)} $;
 Thus
 $ i + ( l+1) r \equiv   j + ( l+1) r   \mod    \bar{m} $, that is  
$ i\equiv j  \mod    \bar{m}$. This  however implies
$i=j$ since  $i < j < k \leq \bar{m}$: a contradiction.  Therefore $s\notin Y$.\\
Finally, assume $s= w_l\in W$.   By construction,   $ \pi(c_i)[t] =  w_l$ implies that  
$  t =  \bar{m} ( \bar{m} -1)  - \lfloor  \frac{\bar{n}}{2} \rfloor -  (i-1) + l- 2$.
On the other hand, $ \pi(c_j)[t] =  w_l$ implies
 by construction that 
$  t =  \bar{m} ( \bar{m} -1) - \lfloor  \frac{\bar{n}}{2} \rfloor -  (j-1) + l- 2$.
As a consequence,
 $ \pi(c_i)[t] =  \pi(c_i)[t] =  w_l$ implies $i=j$, a contradiction. Therefore $s\notin W$.\\
Summarizing, $s\notin U\cup V \cup Y \cup Z \cup W$: a contradiction.

\end{proof}
\ \\ 

%We can now state the proof of Theorem \ref{teo:moveLBsimpHete}:

%\begin{proof} (of Theorem \ref{teo:moveLBsimpHete})\\
Given $ n \geq 36$ and $\frac{n}{6} -2  \geq k \geq 4$, let  $\vec{G}_R $
        be the simple  graph of a {\tt SiHe} system with those values.
         By Claims \ref{lm:SimpleHetero1} and
\ref{lm:SimpleHetero2}, in  the  {\tt SiHe}  system there is 
a simple route $\pi(c_0)$ of period  $q =  \bar{m}^2 -  \bar{m} +  k-1$,
and $k-1$ simple routes ($\pi(c_i), 0<i<k$) of
period  $p= q+1$.
Each  $\pi(c_i)$ with $i>0$
has a distinguished site,  $u_i$, not present in any other route;
furthermore, $\pi(c_i)$ has no connection with $\pi(c_j)$ for $i\neq j$,
while it has  a unique meeting point, $z_i$,
with $\pi(c_0)$.

 Let \al start at $x_0$ at time $t=0$ with $c_0$.
 Since $u_i$ is only in route $\pi(c_i)$, and all
$u_i$'s must be visited,
\al must hitch a ride on all  $c_i$'s.

Let  $t_i$ be the first time \al 
hitches a ride on $c_i$  at
$z_i$.
Notice that once \al is hitching a ride on  carrier $c_i$,
since route $\pi(c_i)$ has no connection with  $\pi(c_j)$, $i\neq j>0$,
to hitch a ride on $c_j$
 \al must
first return at $z_i$ and hitch a ride on $c_0$.
Since $p$ and $(p-1)$ are coprime,
this can happen only at a time $t'> t_i$ such that
 $t' \equiv t_i  \mod(q r)$; that is,
 after at least $p (p-1)$ moves since \al hitched a ride on $c_i$.
 
Since \al must go on all routes (to visit the $u_i's$),
at least $(k-2) p (p-1)$ moves must be performed until
\al hitches a ride on the last carrier, say $c_l$;
then, once the last
distinguished site $z_l$ has been visited,  after
at least $p -  (k -1)$ moves, \al can stop.
Hence the total number of moves  is
at least $ (k-2)\  p\  (p-1) + p -  k + 1 > (k-3)\  p^2$
since 
$p > k$.

 Recall that   $\bar{m} $ is the largest prime number smaller than
    $\frac{1}{2}(n-3k-4)$; since $k \leq \frac{n}{6}-2$,  we have
%$\frac{1}{2}(n-3k-4)  \geq  $
$\bar{m} \geq \frac{1}{4}(n-3k-4) > \frac{n}{2}$;
thus 
\begin{center}
$ p =  \bar{m}^2 -  \bar{m} +  k >  \frac{n^2}{4} -  \frac{1}{2}(n-3k-4) + k >
\frac{1}{4} (n^2 - 2n)$
\end{center}

Hence the total number of moves is more than 

\begin{center}
$(k-3)\  p^2 >  \frac{1}{16} (k-3 ) (n^2 - 2n)^2 = \Omega(k n^4)$
\end{center}
\noindent completing the proof.
\end{proof}
\ \\ 

\begin{proof} {\bf (of Theorem \ref{teo:moveLBirrHetero})}\\
%CONSTRUCTION 3: {"Hetero Star")
Consider a system where $S= \{x_0,  \ldots, x_{q-2}, y_1, \ldots, y_{r-1}, z_1, \ldots z_{k-1}\}$,
where $r<q$, and $q$ and $r$ are coprime,
 $C= \{c_0, c_1 \ldots, c_{k-1}\}$, and the set of routes is defined as follows:

  $$\pi(c_i) =  \left\{\begin{array}{ll}
                     < x_0, y_1, y_2, \ldots, y_{r-1}>&\textrm{for $i=0$}\\
                       <  \alpha(i),  \beta(i),  z_{i} > &\textrm{for $1 \leq  i <  k$}
                     \end{array}
                     \right.
$$

\noindent where  $\alpha(j) = x_{j}, x_{j+1},\ldots, x_{q-2}$, and
 $\beta(j) =x_0,  \ldots,  x_{j-1}$.
In other words, in  the system  there is a irredundant route of period $r$,   $\pi(c)$,
and   $k-1$ irredundant routes of period $q$,  $\pi(c_i)$  for $1 \leq i <  k$. Each of the latter
has a distinguished site (the $z_i$'s), not present in any other route;
furthermore, $\pi(c_i)$ has no connection with $\pi(c_j)$ for $i\neq j$.
On the other hand,  each route $\pi(c_i)$
 has  the same meeting point, $x_0$,
with $\pi(c)$.
Let $t_i$ denote the first time $c$ and $c_i$ meet at $x_0$; 
%{\bf MUST PROVE THAT
%ALL THESE $t_i$ ARE DISTINCT MOD ...}
notice that  if $i\neq j$ then $t_i \not\equiv  t_j \mod(q)$.
Further note that since   $r$ and $q$ are coprime,  $c$  will meet  $c_i$     at time $t$ if
and only if
 $t \equiv t_i  \mod(q\ r)$.

 Let \al start at $x_0$ at time $t=0$ with $c$.
 Since $z_i$ is only in route $\pi(c_i)$, and all
$z_i$'s must be visited,
\al must hitch a ride on all  $c_i$'s.
Notice that once \al is hitching a ride on  carrier $c_i$,
since route $\pi(c_i)$ has no connection with  $\pi(c_j)$, $i\neq j$,
to hitch a ride on $c_j$
 \al must
first return at $x_0$ and hitch a ride on $c$.
To hitch a ride on $c_i$,\al must have been on $c$ at
$x_0$ at some time  $t' \equiv t_i  \mod(q r)$;
 hitching again a ride on $c$ at $x_0$ can happen
only at a time $t' < t'' \equiv t_i  \mod(q r)$; in other words,
after at least $q r$ moves since \al hitched a ride on $c_i$.
Once on $c$ again, to hitch a ride on $c_j$ \al must continue to move
until it reaches $x_0$ at time $t'' < t''' \equiv t_j  \mod(q r)$,
requiring at least $r$ moves.
In other words, to move from a route $\pi(c_i)$ to a different route
$\pi(c_j)$ \al must perform at least $p r + r$ moves.
Since \al must go on all routes (to visit the $y_i's$),
at least $(k-2) (p r + r)$ moves must be performed until
\al hitches a ride on the last carrier, say $c_l$;
then, once the last
distinguished site $z_l$ has been visited after $q$ moves, \al can avoid returning to $a_0$
and stop.
Since at time t=0, $x$ is on $x_0$ and no other carrier is there at that
time, at least $\min{t_i}+1  \geq  r$ moves are performed by \al before it hitches
its first ride on one of the $c_i$'s.
Hence the total number of moves is
at least
\begin{equation}
\label{eq:irr}
  (k-2) (p r + r) + r + p
 \end{equation}

We now have to show how to
use this facts to prove our theorem  for any $n$ and  $k \leq \epsilon\ n$ $(0< \epsilon<1)$.
We will consider two cases, depending on
whether or not  $n-k$ is even.
 Let   $n-k$ be even; if we
 choose $r = \frac{n-k}{2} +1$ and $q=\frac{n-k}{2} +2$,
 then  $n= k + q + r - 3$, and
$r$ and $k$ are coprime; hence  the total number of moves is  that of Expression (\ref{eq:irr}).
%at least
%$  (k-2) (p r + r) + r + p   >  (k-2)  p  r$.
 Since  $k \leq  \epsilon\ n$, then
$n-k \geq (1- \epsilon) n$; thus
\begin{center}
$p\ r =  (\frac{n-k}{2} +1) (\frac{n-k}{2} +2) =
(\frac{(1- \epsilon) n}{2} +1) (\frac{(1- \epsilon) n}{2} +2)$
 \end{center}
Let   $n-k$ be odd; if we
 choose
  $r = \frac{n-k+3}{2} -1$ and $q=\frac{n-k-3}{2} +1$,
 then  $n= k + q + r - 3$, and
$r$ and $k$ are coprime. Hence
the total number of moves is  that of Expression (\ref{eq:irr}).
 Since  $k \leq  \epsilon\ n$, then
$n-k \geq (1- \epsilon) n$;  it follows that
\begin{center}
$p\ r =  (\frac{n-k+3}{2} -1) (\frac{n-k+3}{2} +1) =
(\frac{ (1- \epsilon) n+3}{2} -1) (\frac{ (1- \epsilon) n+3}{2} +1) $
 \end{center}
 That is,  regardless of
 whether $n-k$ is even or odd,
 $p\ r   >  ( \frac{ (1- \epsilon) n}{2})^2$.
 Hence the total number of moves is more than
 \begin{equation}
 \label{eq:irrEven}
(k-2)\ p\ r  >  \frac{1}{4} (1- \epsilon)^2  (k-2)\  n^2.
 \end{equation}
 and
 the theorem holds.

\end{proof}

\end{document}